\documentclass[longauth]{aa}  
\usepackage{ulem}
\usepackage{multirow}
\usepackage{graphicx}
\usepackage{txfonts}
\usepackage[colorlinks=true,linkcolor=red,anchorcolor=blue,citecolor=blue,filecolor=black,menucolor=black,runcolor=black,urlcolor=blue]{hyperref}
\usepackage{float}

\usepackage[dvipsnames]{xcolor}

\begin{document}

   \title{PRISMS. U37126, a very blue, ISM-naked starburst at $z=10.255$ with nearly 100\% Lyman continuum escape fraction}

   \author{R. Marques-Chaves\inst{\ref{inst:Geneva}}\thanks{Rui.MarquesCoelhoChaves@unige.ch}  
   \and J.~\'Alvarez-M\'arquez\inst{\ref{inst:CAB}} 
   \and L.~Colina\inst{\ref{inst:CAB}} 
   \and S.~Kendrew\inst{\ref{inst:ESA-BALTIMORE}} 
   \and Abdurro’uf\inst{\ref{inst:Indiana}}
   \and C.~Blanco-Prieto\inst{\ref{inst:CAB},\ref{inst:UCM}} 
   \and L.~A.~Boogaard\inst{\ref{inst:Leiden}} 
   \and M.~Castellano\inst{\ref{inst:INAF-ROME}}
   \and K.~I.~Caputi\inst{\ref{inst:Kapteyn}}
   \and A.~Crespo-G\'omez\inst{\ref{inst:Baltimore}}
   \and A.~Fontana\inst{\ref{inst:INAF-ROME}}
   \and Y.~Fudamoto\inst{\ref{inst:CHIBA}}
   \and S.~Fujimoto\inst{\ref{inst:Toronto},\ref{inst:Dunlap}}
   \and M.~García-Marín\inst{\ref{inst:ESA-BALTIMORE}}
   \and Y.~Harikane\inst{\ref{inst:Tokyo}}
   \and S.~Harish\inst{\ref{inst:Baltimore}}
   \and T.~Hashimoto\inst{\ref{inst:Tsukuba},\ref{inst:TCHoU}}
   \and T.~Hsiao\inst{\ref{inst:Austin}}
   \and E.~Iani\inst{\ref{inst:ISTA}}
   \and A.~K.~Inoue\inst{\ref{inst:SASE},\ref{inst:Waseda}}
   \and D.~Langeroodi\inst{\ref{inst:DARK}}
   \and R.~Lin\inst{\ref{inst:Amherst}}
   \and J.~Melinder\inst{\ref{inst:Stockholm}}
   \and L.~Napolitano\inst{\ref{inst:INAF-ROME}}
   \and G.~${\rm \ddot{O}}$stlin\inst{\ref{inst:Stockholm}}
   \and P.~G.~Pérez-González\inst{\ref{inst:CAB}} 
   \and C.~Prieto-Jiménez\inst{\ref{inst:CAB},\ref{inst:UCM}} 
   \and P.~Rinaldi\inst{\ref{inst:AURA}}
   \and B.~Rodríguez~Del~Pino\inst{\ref{inst:CAB}}
   \and P.~Santini\inst{\ref{inst:INAF-ROME}}
   \and Y.~Sugahara\inst{\ref{inst:SASE},\ref{inst:Waseda}}
   \and A.~Varo-O'Ferrall\inst{\ref{inst:CAB},\ref{inst:UCM}} 
   \and G.~Wright\inst{\ref{inst:Edinburgh}}
   \and J.~Zavala\inst{\ref{inst:Amherst}} }

   \institute{Geneva Observatory, Department of Astronomy, University of Geneva, Chemin Pegasi 51, CH-1290 Versoix, Switzerland \label{inst:Geneva}
   \and Centro de Astrobiolog\'{\i}a (CAB), CSIC-INTA, Ctra. de Ajalvir km 4, Torrej\'on de Ardoz, E-28850, Madrid, Spain\label{inst:CAB}
   \and European Space Agency (ESA), ESA Office, Space Telescope Science Institute, 3700 San Martin Drive, Baltimore, MD 21218, USA\label{inst:ESA-BALTIMORE}
   \and Department of Astronomy, Indiana University,727 East Third Street, Bloomington, IN 47405, USA\label{inst:Indiana}
   \and Departamento de Física de la Tierra y Astrof\'isica, Facultad de Ciencias F\'isicas, Universidad Complutense de Madrid, E-28040 Madrid, Spain \label{inst:UCM}
   \and Leiden Observatory, Leiden University, PO Box 9513, NL-2300 RA Leiden, The Netherlands\label{inst:Leiden}
   \and INAF – Osservatorio Astronomico di Roma, via Frascati 33, 00078, Monteporzio Catone, Italy\label{inst:INAF-ROME}  
   \and Kapteyn Astronomical Institute, University of Groningen, P.O. Box 800, 9700AV Groningen, The Netherlands\label{inst:Kapteyn}
   \and Space Telescope Science Institute (STScI), 3700 San martin Drive, Baltimore, MD 21218, USA\label{inst:Baltimore}
   \and Center for Frontier Science, Chiba University, 1-33 Yayoi-cho, Inage-ku, Chiba 263-8522, Japan\label{inst:CHIBA}
   \and David A. Dunlap Department of Astronomy and Astrophysics, University of Toronto, 50 St. George Street, Toronto, Ontario, M5S 3H4, Canada\label{inst:Toronto}
   \and Dunlap Institute for Astronomy and Astrophysics, 50 St. George Street, Toronto, Ontario, M5S 3H4, Canada\label{inst:Dunlap}
   \and Institute for Cosmic Ray Research, The University of Tokyo, 5-1-5 Kashiwanoha, Kashiwa, Chiba 277-8582, Japan\label{inst:Tokyo}
   \and Division of Physics, Faculty of Pure and Applied Sciences, University of Tsukuba, Tsukuba, Ibaraki 305-8571, Japan\label{inst:Tsukuba}
   \and Tomonaga Center for the History of the Universe (TCHoU), Faculty of Pure and Applied Sciences, University of Tsukuba, Tsukuba, Ibaraki 305-8571, Japan\label{inst:TCHoU}   
   \and Department of Astronomy, University of Texas, Austin, TX 78712, USA\label{inst:Austin}
   \and Institute of Science and Technology Austria (ISTA), Am Campus 1, 3400 Klosterneuburg, Austria\label{inst:ISTA}
   \and Department of Physics, School of Advanced Science and Engineering, Faculty of Science and Engineering, Waseda University, 3-4-1 Okubo, Shinjuku, Tokyo 169-8555, Japan \label{inst:SASE}
   \and Waseda Research Institute for Science and Engineering, Faculty of Science and Engineering, Waseda University, 3-4-1 Okubo, Shinjuku, Tokyo 169-8555, Japan \label{inst:Waseda}
   \and DARK, Niels Bohr Institute, University of Copenhagen, Jagtvej 155A, 2200 Copenhagen, Denmark\label{inst:DARK}
   \and University of Massachusetts Amherst, 710 North Pleasant Street, Amherst, MA 01003-9305, USA\label{inst:Amherst}
   \and Department of Astronomy, Stockholm University, Oscar Klein Centre, AlbaNova University Centre, 106 91 Stockholm, Sweden\label{inst:Stockholm}
   \and AURA for the European Space Agency (ESA), Space Telescope Science Institute, 3700 San Martin Dr., Baltimore, MD 21218, USA \label{inst:AURA}
   \and UK Astronomy Technology Centre, Royal Observatory Edinburgh, Blackford Hill, Edinburgh EH9 3HJ, UK\label{inst:Edinburgh}
   }

   \date{Received ; accepted}

\abstract
{We present very deep ($\approx$\,11\,hours on-source) JWST/MIRI low-resolution spectroscopy of the rest-frame optical emission of U37126, a UV-bright ($M_{\rm UV}\simeq -20$), mildly lensed ($\mu\simeq 2.2$) galaxy at $z = 10.255$. The continuum emission is well detected in both NIRSpec and MIRI spectra, yet no nebular recombination or metal emission lines are observed ($EW_{0}\,(\rm H\beta+$[O\,{\sc iii}]$)\leq 300$\,\AA{ }and $EW_{0}\,(\rm H\alpha)\leq 400$\,\AA, at 3$\sigma$). 
Combined with the exceptionally blue UV continuum slope, $\beta_{\rm UV} \simeq -2.9$, and flat Balmer break, these constraints indicate a stellar population dominated by very young and massive stars with a strongly suppressed nebular contribution. Comparisons with synthetic stellar population models indicate that U37126 requires both a very high ionizing photon production efficiency, log($\xi_{\rm ion} / \rm Hz\, erg^{-1}) \simeq 25.75$, and a nearly unit LyC escape fraction, of $f_{\rm esc} \geq 86\%$ (3$\sigma$) based on H$\alpha$ flux limit and $f_{\rm esc} = 0.94 \pm 0.06$ derived independently from SED fitting.
The best-fit SED yields a (de-lensed) stellar mass of $M_{\star} \simeq 10^{7.8}\,M_{\odot}$ and a star-formation rate of $\mathrm{SFR} \simeq 10\,M_{\odot}\,\mathrm{yr^{-1}}$ ($\rm sSFR \sim 160$\,Gyr$^{-1}$), that along with its very compact size, $r_{\rm eff}\simeq 61$\,pc, yields very high stellar mass and star-formation-rate surface densities, $\Sigma_{M\star} \simeq 3 \times 10^{3} M_{\odot}\,\rm pc^{-2}$ and $\Sigma_{\rm SFR} \simeq 400\, M_{\odot}\,\rm yr^{-1}\, kpc^{-2}$. Together with the lack of detectable nebular emission, these properties suggest that U37126 is undergoing an ``ISM-naked'' starburst phase, possibly driven by an extremely efficient gas-to-star conversion followed by strong feedback that has cleared the remaining gas from its stellar core, allowing most LyC photons to escape. Finally, we show that even a small fraction of galaxies like U37126 ($\simeq 3\%$--$6\%$), with extreme LyC production and escape, could contribute disproportionately ($\simeq 50\%$--$100\%$) to the ionizing photon budget during cosmic reionization.}

\titlerunning{A strong LyC emitter at $z\simeq10$}
\maketitle

\section{Introduction}\label{Sect:intro}

The James Webb Space Telescope (JWST) is fundamentally reshaping our understanding of galaxy formation in the early Universe by enabling the detection and detailed characterization of the first galaxies within the first few hundred million years after the Big Bang \citep[e.g.,][]{Castellano2022ApJ...938L..15C,  Curtis-Lake+23, Harikane+23a, Perez-Gonzalez+23b,   Carniani+24, Castellano2024, Harikane2024ApJ...960...56H,  Naidu2025arXiv250511263N, Napolitano2025A&A...693A..50N, Chemerynska2026MNRAS.tmp...12C}. Deep NIRCam imaging and NIRSpec spectroscopy provide unprecedented access to the rest-frame ultraviolet (UV) and optical emission of galaxies at redshifts $z \gtrsim 7$, while MIRI extends such studies to even higher redshifts. Together, JWST observations are now robustly constraining the physical properties of the earliest galaxies that were previously inaccessible, including their stellar populations, star-formation histories, and nebular emission \citep[e.g.,][]{Alvarez-Marquez+23-MACS, Bunker+23, Fujimoto2024ApJ...977..250F, Alvarez2025A&A...695A.250A, Helton2025arXiv251219695H, Roberts-Borsani2025arXiv250821708R, Tang2025arXiv250708245T, Zavala2025NatAs...9..155Z, Donnan2026arXiv260111515D}.

The UV continuum slope, $\beta_{\rm UV}$ ($f_{\rm \lambda} \propto \rm \lambda^{\beta_{\rm UV}}$), is a widely used spectral diagnostic for the physical conditions in star-forming galaxies. Prior to JWST, observations established that typical star-forming galaxies at $z \gtrsim 2$ exhibit relatively blue UV slopes ($\beta \simeq -2$), and that UV continua become systematically bluer toward higher redshifts and fainter UV luminosities \citep[e.g.,][]{Bouwens2012ApJ...754...83B, Finkelstein2012ApJ...756..164F, Bouwens2014ApJ...793..115B, Bhatawdekar2021ApJ...909..144B}. These trends are generally interpreted as reflecting lower dust attenuation and younger stellar populations in early galaxies. Recent JWST observations have extended these measurements even above $z > 10$ with improved precision, revealing an increasing prevalence of very steep UV slopes among the earliest galaxies  \citep[e.g.,][]{Topping2022ApJ...941..153T, Cullen2023MNRAS.520...14C, Cullen2024MNRAS.531..997C,  Morales2024ApJ...964L..24M, Saxena2024arXiv241114532S, Topping2024MNRAS.529.4087T, Dottorini2025A&A...698A.234D, Messa2025A&A...694A..59M}.

While moderately blue $\beta_{\rm UV}$ can be readily produced by hot, massive stars expected in young stellar populations, achieving very steep slopes ($\beta \lesssim -2.6$) is considerably more challenging and requires additional physical conditions. At the young ages ($\lesssim 10$\,Myr) 
necessary to generate such intrinsically blue stellar continua, nebular emission powered by ionizing photons typically contributes significantly to the observed spectrum, acting to redden the emergent UV spectrum \citep[e.g.,][]{Bouwens2010ApJ...708L..69B}. As a result, stellar population models generally predict that UV slopes approaching $\beta \simeq -3$ should be rare or unobservable in systems where ionizing photons are efficiently reprocessed by surrounding gas \citep[e.g.,][]{Katz2025OJAp....8E.104K}.
Extremely blue UV slopes thus point to conditions in which the contribution of nebular emission to the emergent UV spectrum is strongly suppressed. This naturally occurs when a significant fraction of Lyman continuum (LyC, with $>13.6$\,eV) ionizing photons escape from H\,{\sc ii} regions before being reprocessed into nebular line and continuum emission. 

Galaxies exhibiting extremely steep UV slopes together with weak nebular emission thus represent compelling candidates for systems with exceptionally high LyC escape fractions and enhanced ionizing photon output. This framework was first explored by \cite{Zackrisson2013, Zackrisson2017ApJ...836...78Z} who showed that the combination of steep UV slopes and weak rest-optical emission lines provides a powerful means of identifying strong LyC emitters. 
This not only has the advantage of selecting sources with very high LyC $f_{\rm esc}$, but also with exceptionally high ionizing photon production efficiencies, $\xi_{\rm ion}= Q_{\rm H}/L_{\rm UV}$, where $Q_{\rm H}$ is the hydrogen-ionizing photon production rate and $L_{\rm UV}$ is the UV luminosity. Such high $\xi_{\rm ion}$ values are expected in very young stellar populations required to reproduce extremely steep UV continuum slopes.

Nevertheless, and with a few exceptions \citep{Marques-Chaves2022MNRAS.517.2972M, Kim2023ApJ...955L..17K}, confirmed low-redshift LyC emitters exhibit only moderately blue $\beta_{\rm UV}$ \citep{Chisholm2022} and generally strong nebular emission lines, with rest-frame H$\beta$ equivalent widths exceeding $>150$\,\AA~\citep[e.g.,][]{Izotov2016MNRAS.461.3683I, Izotov2018MNRAS.478.4851I, Flury2022ApJ...930..126F}. Only recently, JWST started to reveal a rare, yet non-negligible population of $z>6$ sources with both extremely blue UV slopes and weak nebular emission lines \citep{Topping2022ApJ...941..153T, Hainline2024ApJ...976..160H, Donnan2025ApJ...993..224D, Yanagisawa2025ApJ...988...86Y}, often with inferred LyC escape fractions well above $f_{\rm esc} >50\%$ \citep[e.g.,][]{Giovinazzo2025arXiv250701096G}.   

In this work, we present the discovery of another such system, UNCOVER-37126 at $z=10.255$ (hereafter U37126), previously identified by \cite{Atek2023MNRAS.524.5486A} and spectroscopically confirmed by \cite{Fujimoto2024ApJ...977..250F}. Leveraging ultra-deep JWST/MIRI spectroscopy of its rest-frame optical emission, together with ancillary NIRCam imaging and NIRSpec spectroscopy, we show that U37126 exhibits an extremely steep UV continuum slope and a non-detection of nebular emission lines, consistent with LyC escape fraction close to unity. The paper is organized as follows. Section~\ref{Sect2:data_calibration_lines} presents the MIRI, NIRCam, and NIRSpec observations. Section~\ref{Sect:results_dis}  describes the observational results, which are discussed in Section~\ref{Sect4:disc}. Finally, Section~\ref{Sect:conclusion_Summary} summarizes our conclusions. Throughout this work, we use a concordance cosmology with $\Omega_\mathrm{m}$\,=\,0.31, and H$_0$\,=\,67.7\,km\,s$^{-1}$\,Mpc$^{-1}$ \citep{PlanckCollaboration18VI}. 

\section{JWST Observations and data reduction}\label{Sect2:data_calibration_lines}

U37126 was observed with the Low Resolution Spectrograph (LRS, \citealt{Kendrew2015PASP..127..623K}) of the Mid-InfraRed Instrument (MIRI, \citealt{Rieke+15, Wright+15, Wright+23}) on 7-8 November 2025 as part of the "PRImordial galaxy Survey with MIRI Spectroscopy at $z\sim 10$" (PRISMS, program ID 8051; PIs: J. Álvarez-Márquez \& L. Colina; \citealt{alvarez-marquez2026arXiv260202323A}). These observations provide a spectral coverage between $4.80-14.0\,\mu$m wavelength range using a $0\farcs51 \times 4\farcs7$ slit, i.e., from $\lambda_{\rm rest} \simeq 0.43-1.20\,\mu$m in the rest-frame, and a spectral resolution of $R\sim 100$. The total on-source integration time was 39,696\,s ($\simeq 11.0$\,h), obtained using a customized four-point dither strategy repeated over 12 dithers. The target acquisition was done using the F560W filter on a GAIA DR3 reference star.

The data reduction follows \cite{alvarez-marquez2026arXiv260202323A}. Briefly, we use version 1.20.2 of the JWST calibration pipeline and CRDS context \texttt{jwst\_1464.pmap}, following the standard MIRI LRS procedures \citep{bushouse_2025_17515973} with additional custom steps to optimize background subtraction and artifact removal. These include wavelength masking, master and residual background subtraction, and sigma clipping to mitigate detector artifacts and cosmic-ray residuals. 
The final combined 2D spectrum and 1D extracted spectrum were produced using the pipeline Stage 3, with the 1D spectrum extracted using a $0\farcs44$ aperture and corrected for aperture losses using the standard JWST reference files (\texttt{jwst\_miri\_apcorr\_0017.fits}). For more details, see \cite{alvarez-marquez2026arXiv260202323A}.

In addition to the MIRI LRS observations, NIRCam imaging and NIRSpec spectroscopy of U37126 are publicly available as part of the UNCOVER project\footnote{\url{https://jwst-uncover.github.io/}}. NIRCam photometry, combining medium- and broad-band filters from F070W to F480M, is taken from the UNCOVER DR3 SUPER catalog \citep{Suess2024ApJ...976..101S, Weaver2024ApJS..270....7W}. NIRSpec Micro-Shutter Assembly (MSA) spectroscopy, obtained with the low-resolution PRISM ($R\sim100$), is taken from the UNCOVER data release 4 \citep{Bezanson2024ApJ...974...92B}. We use the fully reduced and calibrated NIRSpec spectrum ($\simeq 4.4$\,hours on-source) published by \citet{Fujimoto2024ApJ...977..250F}. Finally, we adopt the updated gravitational magnification from the UNCOVER DR4, $\mu = 2.19 \pm 0.05$ \citep{Furtak2023MNRAS.523.4568F, Price2025ApJ...982...51P}.

\begin{figure*}
  \centering
  \includegraphics[width=0.98\textwidth]{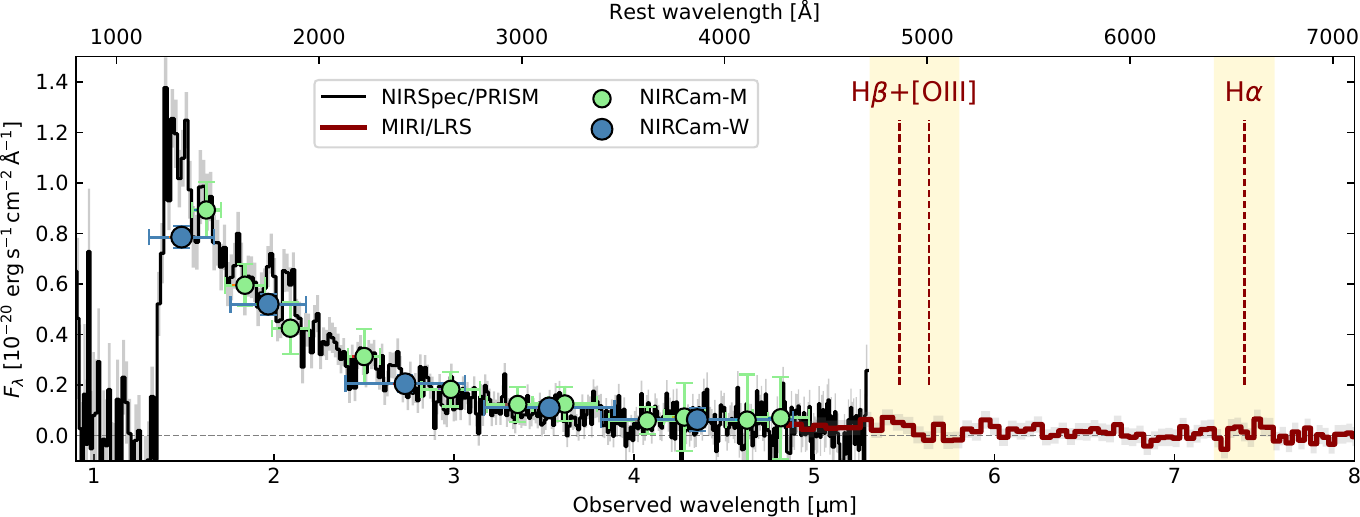}
  \caption{NIRSpec/PRISM (black) and MIRI/LRS (red) spectra of U37126 with the 1$\sigma$ uncertainties shown in grey. The expected locations of the main rest-frame optical emission lines, H$\beta$, [O\,{\sc iii}]\,$\lambda\lambda$4960,5008, and H$\alpha$, are indicated with red dashed lines. NIRCam photometry from broad- and medium-band filters is overplotted in blue and green, respectively.}
  \label{fig_spectrum}
\end{figure*}

To place all JWST observations of U37126 on a consistent relative flux scale, we first compute synthetic NIRSpec photometry convolving the NIRSpec spectrum with the filter transmission functions of all medium- and broad-band NIRCam filters, where the source is significantly detected ($\geq 3\sigma$; F150W–F480M) and compare it to the corresponding NIRCam photometry. This yields a scaling factor of $\simeq 1.41$, which is applied to the NIRSpec spectrum. For the MIRI LRS spectrum, we compute synthetic photometry over the common spectral range $\lambda_{\rm obs} = 4.85-5.30\,\mu$m, where the continuum is significantly detected in both NIRSpec and MIRI. We find consistent flux densities between the two spectra in this region, and therefore apply no additional scaling to the MIRI LRS spectrum. Figure~\ref{fig_spectrum} shows the combined NIRSpec and MIRI spectra together with the NIRCam photometry.

\section{Results}\label{Sect:results_dis}

U37126 was first identified by \cite{Atek2023MNRAS.524.5486A} as a photometric redshift candidate at $z_{\rm phot} = 10.60^{+0.21}_{-0.31}$ based on NIRCam photometry (ID: 39074). A spectroscopic redshift of $z_{\rm spec} = 10.255 \pm 0.001$ was subsequently reported by \cite{Fujimoto2024ApJ...977..250F}, based on the unambiguous detection of the Ly$\alpha$ break. A tentative detection of the N\,{\sc iii}] $\lambda$1750 emission line was also discussed in \cite{Fujimoto2024ApJ...977..250F}; however, we do not find significant emission at the reported wavelength, nor evidence for any other rest-frame UV or optical emission lines in the NIRSpec and MIRI spectra (Fig.~\ref{fig_spectrum}). We therefore adopt the redshift $z = 10.255 \pm 0.001$ inferred from the Ly$\alpha$ break throughout this work.
Finally, U37126 was recently observed in the far-infrared with ALMA; however, neither dust continuum emission nor [O,{\sc iii}] 88\,$\mu$m line emission was detected, yielding upper limits of $\log(M_{\rm dust}/M_{\odot}) < 6.06$ and $L({\rm [O\,III]}\,88\,\mu{\rm m}) < 2 \times 10^{8}\,L_{\odot}$, respectively \citep{Algera2025arXiv251214486A}.

\subsection{Spectral properties from NIRSpec, NIRCam, and MIRI}\label{spectral_properties}

The $\beta_{\rm UV}$ is measured using several independent methods based on both NIRSpec and NIRCam.
We first fit the NIRSpec/PRISM spectrum over a broad and continuous rest-frame window, $1400-3600$\,\AA, thereby excluding regions potentially affected by intergalactic medium (IGM) absorption blueward of $1400$\,\AA\ and by the Balmer break at longer wavelengths. A power-law fit using \texttt{lmfit} \citep{Newville2025zndo..15014437N} yields $\beta_{\rm UV} = -2.79 \pm 0.07$.
We then adopt the continuum windows defined by \citet{Calzetti1994ApJ...429..582C}, which avoid ISM absorption features, stellar P-Cygni profiles, and nebular emission lines. To remain consistent with the above criteria, we restrict the fit to windows at $\lambda_{\rm rest} \geq 1400$\,\AA, obtaining $\beta_{\rm UV} = -2.83 \pm 0.09$.
As an alternative spectral diagnostic, we fit a first-order polynomial to the wavelength intervals $2180-2220$\,\AA\ and $2780-2820$\,\AA\ following \citet{Leitherer+99}. This method yields $\beta_{\rm UV} = -3.05 \pm 0.16$.
Finally, we use the NIRCam broad-band photometry in F200W and F277W, which sample the rest-frame range $\lambda_{\rm rest} \simeq 1800-2500$\,\AA. From this, we obtain $\beta_{\rm UV} = -2.86 \pm 0.20$.
Taken together, these independent diagnostics consistently indicate that U37126 shows a steep UV continuum slope, all with $\beta_{\rm UV} \lesssim -2.7$. Taking the simple average of these measurements, obtained from different spectral regions and instruments (NIRSpec and NIRCam), and adopting their standard deviation as the uncertainty, we derive a fiducial UV slope of $\beta_{\rm UV} = -2.88 \pm 0.10$.

Another key spectral diagnostic is the strength of the Balmer break, which is sensitive to the age of the stellar population and to the relative contribution of nebular emission. For consistency and to facilitate a direct comparison with the stellar population models discussed next in Section~\ref{predictions}, we estimate the Balmer break strength of U37126 from the NIRSpec spectrum as the flux density ratio (in $F_{\nu}$) between $\lambda_{\rm rest} = 4200$\,\AA\ and $\lambda_{\rm rest} = 3400$\,\AA, measured within spectral windows of width $\Delta\lambda_{\rm rest} = 400$\,\AA. We find a ratio of $F_{\nu}^{4200\AA}/F_{\nu}^{3400\AA} = 0.86 \pm 0.21$.
For comparison, using the observed NIRCam photometry in the F356W and F444W filters, which sample the emission shortward and longward of the Balmer break, respectively, we derive a consistent flux density ratio of $F_{\nu}^{\rm F444W}/F_{\nu}^{\rm F356W} = 0.88 \pm 0.08$.
As discussed further in Section~\ref{predictions}, the inferred Balmer break strength rules out single-burst stellar populations older than $\gtrsim 6$\,Myr, as well as continuous star formation histories with ages $\gtrsim 16$\,Myr.

Finally, we analyze the MIRI/LRS spectrum. At the redshift of U37126 ($z=10.255$), the H$\beta$, [O\,{\sc iii}]\,$\lambda\lambda$\,4960,5008, and H$\alpha$ emission lines are expected at $\lambda_{\rm obs}=5.47\,\mu$m, $5.64\,\mu$m, and $7.39\,\mu$m, respectively, yet none of them are significantly detected (Fig.~\ref{fig_spectrum}).\footnote{The reduced NIRSpec spectrum (v4) from the Dawn JWST Archive extends up to $\lambda_{\rm obs} \simeq 5.5\,\mu$m, but H$\beta$ is also not detected.} We estimate their flux upper limits by measuring the root mean square (rms) of the spectrum within a rest-frame window of $\Delta\lambda=300$\,\AA\ centered on the expected position of each line (yellow regions in Figure~\ref{fig_spectrum}), and adopt the MIRI/LRS instrumental resolutions of $R\simeq 49$ and $R\simeq 89$ for [O\,{\sc iii}] and H$\alpha$, respectively \citep{Kendrew2015PASP..127..623K}, i.e., assuming that the emission lines are unresolved. We also consider an intrinsic [O\,{\sc iii}] 5008/4960 ratio of 2.98 \citep{Osterbrock+Ferland+06}.

Under these assumptions, we derive $3\sigma$ limits of $F\,(\rm [OIII]\,\lambda 5008) \leq 6.6 \times 10^{-19}$ erg\,s$^{-1}$\,cm$^{-2}$ and $F\,(\rm H\alpha) \leq 6.9 \times 10^{-19}$\,erg\,s$^{-1}$\,cm$^{-2}$. The continuum emission is significantly detected in the MIRI/LRS ranges $\lambda_{\rm obs}=4.8-5.6\,\mu$m ($37.2 \pm 4.1$\,nJy) and $5.85-6.73\,\mu$m ($26.1 \pm 5.3$\,nJy), while no significant signal is observed at $\lambda_{\rm obs}\geq 7.0\,\mu$m ($\leq 28$\,nJy at $3\sigma$). We estimate the continuum emission at the wavelengths of [O\,{\sc iii}]\,$\lambda5008$ and H$\alpha$ from the best-fit SED (Section~\ref{SED}), finding continuum flux densities of $\simeq 35.8$\,nJy and $\simeq 27.7$\,nJy, respectively.
These values imply $3\sigma$ rest-frame equivalent-width limits under $EW_{0}({\rm [O\,III]\,\lambda5008}) \leq 174$\,\AA\ and $EW_{0}({\rm H\alpha}) \leq 400$\,\AA. For H$\beta$, we assume case-B recombination with an intrinsic line ratio of $I_{\rm H\alpha}/I_{\rm H\beta} = 2.78$, assuming $T_{e}=1.5\times10^{4}$\,K and $n_{e}=10^{3}$\,cm$^{-3}$ \citep{Luridiana+15}. This yields an upper limit of $F({\rm H}\beta) \leq 2.5 \times 10^{-19}$\,erg\,s$^{-1}$\,cm$^{-2}$ and $EW_{0}({\rm H}\beta) \leq 64$\,\AA. We emphasize that the H$\beta$ flux and equivalent-width limits are dependent on the assumed case-B recombination conditions. These measurements are summarized in Table~\ref{table1}

\begin{figure}
  \centering
  \includegraphics[width=0.45\textwidth]{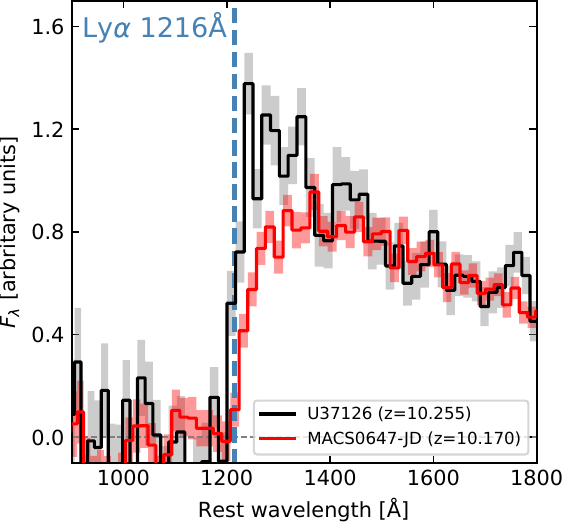}
  \caption{Comparison of the rest-frame Ly$\alpha$ break of U37126 (black) with that of MACS0647-JD ($z=10.170$) which shows a strong damped Ly$\alpha$ absorption feature ($N_{\rm HI} \simeq 2.5 \times 10^{22}$\,cm$^{-2}$, \citealt{Heintz2024Sci...384..890H}). }
  \label{fig_LyA}
\end{figure}

Finally, we briefly examine the spectral region around Ly$\alpha$. Figure~\ref{fig_LyA} shows the rest-frame Ly$\alpha$ break of U37126 and, for comparison, that of MACS0647-JD \citep{Heintz2024Sci...384..890H}, a star-forming galaxy at a similar redshift ($z = 10.170$) and with comparable global properties (e.g., stellar mass and star formation rate). MACS0647-JD exhibits a strong damped Ly$\alpha$ absorption feature, corresponding to a neutral hydrogen column density of $N_{\rm HI} \simeq 2.5 \times 10^{22}$\,cm$^{-2}$ \citep{Heintz2024Sci...384..890H}. In contrast, U37126 shows a very sharp Ly$\alpha$ break, suggestive of low H\,{\sc i} column density along the line of sight. A recent analysis of U37126 by \citet{Mason2026A&A...705A.114M} modeled the Ly$\alpha$ break using the low-resolution PRISM spectrum and inferred a neutral hydrogen column density of $\log(N_{\rm HI}/{\rm cm}^{-2}) = 19.2 \pm 1.5$. Although formally larger than the optically thin limit for LyC escape ($N_{\rm HI} \lesssim 2 \times 10^{17}$\,cm$^{-2}$), the large uncertainty remains statistically consistent with a very high escape fraction at $\sim$1.3$\,\sigma$ level. 

\begin{table}
\begin{center}
\caption{Summary of the properties of U37126. Global quantities have been corrected for lensing magnification. \label{table1}}
\begin{tabular}{l c c}
\hline \hline
\smallskip
\smallskip
Property  & Value & Uncertainty  \\
\hline 
R.A. [J2000]  & 00:14:21.63 & $0.1^{\prime \prime}$ \\
Dec. [J2000]  &  $-$30:21:35.10 & $0.1^{\prime \prime}$ \\
$z$   & 10.255 & $0.001$   \\
$\mu$   & 2.19 & 0.05   \\
$M_{\rm UV}$ [AB]   & $-20.10$ & $0.05$ \\
Age [Myr, CSFH]  & $6.8$   &   $1.6$  \\
SFR [$M_{\odot}$~yr$^{-1}$] & $9.6$ & $4.6$   \\
log($M_{\star}/M_{\odot}$)   & $7.77$ & $0.06$ \\
E(B-V) [mag.]  & $0.01$   &   $0.01$   \\
$r_{\rm eff}$ [pc]  & $61$ & $6$ \\
sSFR [Gyr$^{-1}$]  & 160   & 80  \\ 
$\log(\Sigma_{M\star}$ / $M_{\odot}\, \rm pc^{-2}$) & 3.40 & 0.10  \\
$\log(\Sigma_{\rm SFR}$ / $M_{\odot}\,\rm yr^{-1}\,kpc^{-2}$) & 2.61 & 0.22  \\
$\beta_{\rm UV}$  & $-$2.88 & 0.10 \\
$EW_{0}$\,([O\,{\sc iii}]\,$\lambda$5008) [\AA] & $\leq 174$ & (3$\sigma$) \\
$EW_{0}$\,(H$\beta$) [\AA] & $\leq 64$ & (3$\sigma$) \\
$EW_{0}$\,(H$\alpha$) [\AA] & $\leq 400$ & (3$\sigma$) \\
$L\rm \,([OIII]\,5008\AA)$ [erg\,s$^{-1}$] & $\leq 4.3\times10^{41}$  & (3$\sigma$) \\
$L\rm \,(H\alpha)$ [erg\,s$^{-1}$] & $\leq 4.5\times10^{41}$  & (3$\sigma$) \\
log($\xi_{\rm ion})$ $\rm [Hz\, erg^{-1}]$ &  $25.75$ & $0.09$\\
$f_{\rm esc}$\,(LyC) -- H$\alpha$ & $\geq 0.86$ & (3$\sigma$) \\
$f_{\rm esc}$\,(LyC) -- SED & 0.94 & 0.06 \\
\hline 
\end{tabular}
\end{center}
\end{table}

\subsection{Predictions from synthetic stellar models}\label{predictions}

\begin{figure*}
  \centering
  \includegraphics[width=0.99\textwidth]{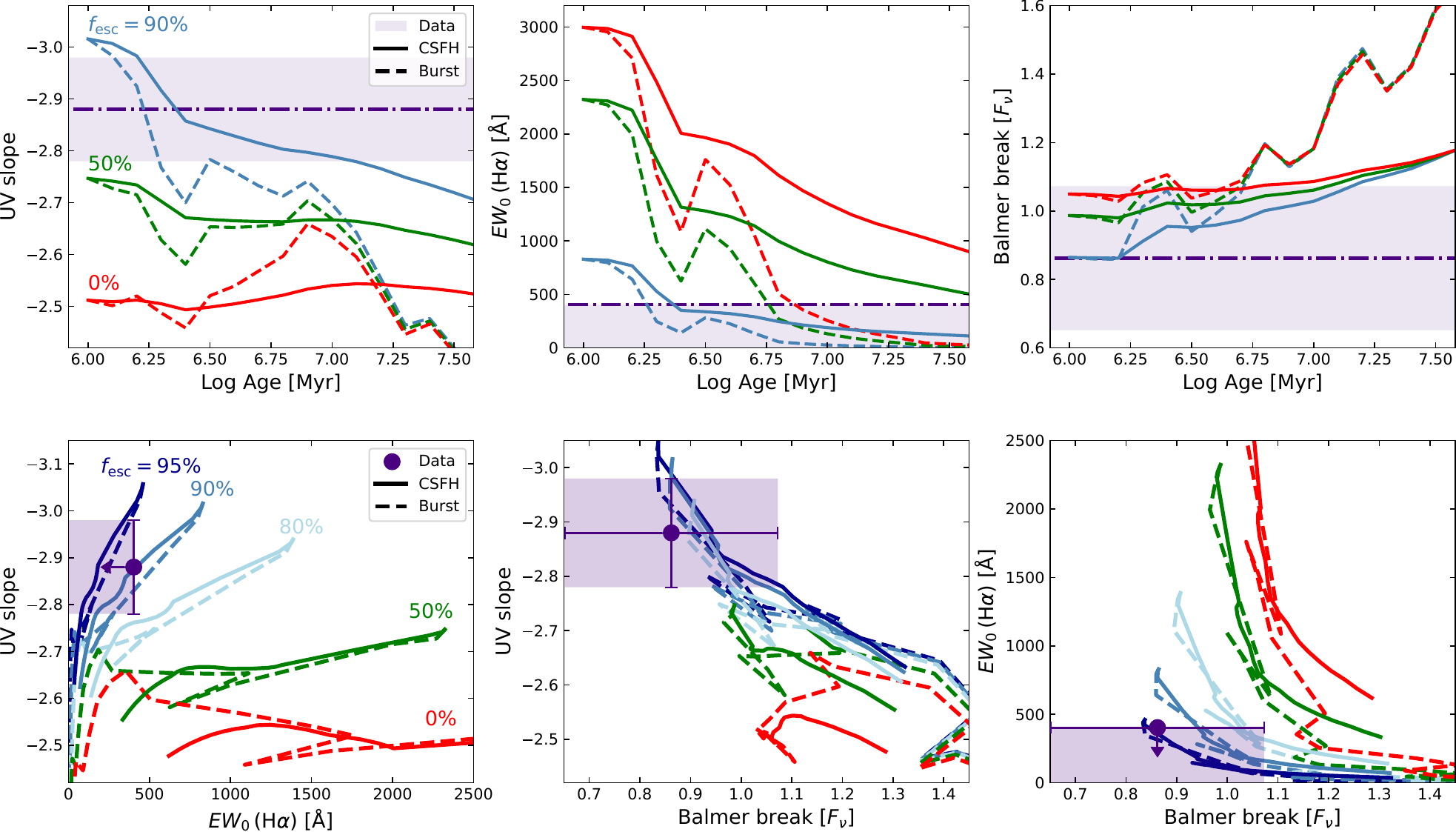}
  \caption{Predictions from BPASS synthetic stellar and nebular emission models ($Z/Z_{\odot}\simeq 0.15$, $T_{e}=1.5\times 10^{4}$\,K, and $n_{e}=10^{3}\,$cm$^{-3}$) for the UV continuum slope ($\beta_{\rm UV}$), $EW_{0}$\,(H$\alpha$), and the Balmer break strength ($F_{\nu}^{4200\AA}/F_{\nu}^{3400\AA}$) as a function of the age (top) and their combination (bottom). Solid and dashed lines correspond to constant star formation (CSFH) and instantaneous burst models (Burst), respectively. Model sequences are shown for LyC escape fractions $f_{\rm esc}$ ranging from 0\% (red) to 90\% (blue); additional models with $f_{\rm esc}=80\%$ and 95\% are shown in the bottom panels only (light and dark blue, respectively). Violet dot-dashed lines (top) and circles (bottom) indicate the observational constraints for U37126, with the shaded regions representing the 1$\sigma$ uncertainties, except for $EW_{0}$\,(H$\alpha$), which represents the $3\sigma$ upper limit.}
  \label{fig_BPASS_predictions}
\end{figure*}

We now compare the observed spectral properties derived in Section~\ref{spectral_properties} with predictions from synthetic stellar population models. We adopt the Binary Population and Spectral Synthesis models (BPASS v2.2.1; \citealt{Stanway+18}), using their default IMF with slope of $-2.35$ and an upper mass cutoff of $300\,\rm M_{\odot}$ (\texttt{imf135\_300}). 
We assume a metallicity $Z=0.003$ ($Z/Z_{\odot}\simeq 0.15$ for $Z_{\odot}=0.02$), consistent with values reported for galaxies of comparable $M_{\rm UV}$ at similar redshifts \citep[e.g.,][]{Stiavelli+23, Boyett2024NatAs...8..657B, Hsiao+2024_MIRI, Alvarez2025A&A...695A.250A, Helton2025arXiv251219695H}. Predictions for additional metallicities, from solar to $Z/Z_{\odot}=0.5\%$, are presented in Appendix~\ref{appendix1} (figures~\ref{fig_appendix_A1} to \ref{fig_appendix_A4}, respectively).

We consider both instantaneous-burst and constant star formation histories, with ages from 1 to 100\,Myr. For each model, we compute the ionizing photon production rate $Q_{\rm H}$ and predict the associated nebular continuum emission using \texttt{PyNeb} \citep{Luridiana+15}. We assume nebular conditions with $T_{e}=1.5\times 10^{4}$\,K and $n_{e}=10^{3}\,$cm$^{-3}$ expected at very high redshifts \citep[e.g.,][]{Isobe+23}, and include free–free and free–bound emission by H and He, and the two-photon continuum of H. Hydrogen recombination-line luminosities (e.g., H$\alpha$) are obtained using the coefficients from \cite{Osterbrock+Ferland+06}. We additionally include the contribution of the H$\gamma$ and H$\delta$ emission lines, as these are sampled by the spectral regions used to infer the Balmer break strength ($F_{\nu}^{4200\AA}/F_{\nu}^{3400\AA}$, see next). The total emergent spectrum is then given by $F_{\rm total} = F_{\rm stellar} + (1-f_{\rm esc}) \times F_{\rm nebular}$, where the factor $(1-f_{\rm esc})$ accounts for the ionizing escape fraction, i.e., LyC photons not reprocessed into nebular emission. We also assume negligible dust attenuation, given the extremely steep UV slope.

For each model and $f_{\rm esc}$ we measure the UV slope, Balmer-break strength, and $EW_{0}\,(\mathrm{H\alpha})$ using the same methodology applied to U37126 (Section~\ref{spectral_properties}). Figure~\ref{fig_BPASS_predictions} (top) shows the predicted quantities as a function of age for $f_{\rm esc}=0$, 0.5, and 0.9 (red, green, blue, respectively). Overall, we recover the expected trends already discussed in previous works: increasing $f_{\rm esc}$ suppresses the nebular continuum and line emission, yielding steeper UV slopes and lower $EW_{0}\,(\mathrm{H\alpha})$ at fixed age \citep[e.g.,][]{Zackrisson2013, Zackrisson2017ApJ...836...78Z}. 
UV slopes steeper than $\beta_{\rm UV}<-2.7$ necessarily require a substantial escape of ionizing photons, and this should hold irrespective of the assumed IMF, metallicity, and age of the underlying stellar population \citep[cf.][]{Katz2025OJAp....8E.104K, Schaerer2025A&A...693A.271S}, and more broadly, for any ionizing source.

The Balmer break, traced here by the flux density ratio $F_{\nu}^{4200\AA}/F_{\nu}^{3400\AA}$, is close to unit for all burst and constant-SFH models at young ages, reflecting the strong contribution of nebular continuum in the $F_{\nu}^{3400\AA}$ but also the emission lines (H$\gamma$ and H$\delta$) to $F_{\nu}^{4200\AA}$, unless $f_{\rm esc}$ is very high ($90\%$, blue). At ages $\gtrsim 6$\,Myr ($\gtrsim 15$\,Myr) for single burst models (constant star formation), the strength of the Balmer break increases with the stellar age, reflecting the rising contribution of less massive stars (A-type) to the integrated spectrum \citep[e.g.,][]{Kuruvanthodi2024A&A...691A.310K, Looser2024Natur.629...53L, Baker2025A&A...697A..90B}.

\begin{figure}
  \centering
  \includegraphics[width=0.45\textwidth]{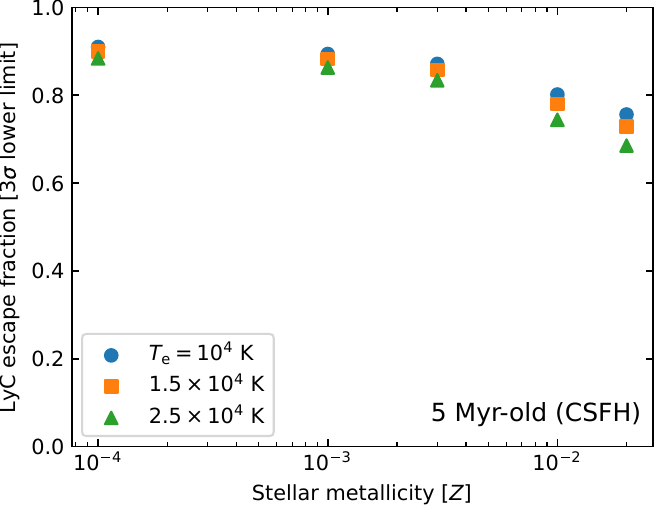}
  \caption{Limits ($3\sigma$) on the Lyman-continuum escape fraction derived from the upper limit on the observed (de-lensed) H$\alpha$ luminosity, as a function of stellar metallicity (from $0.5\%$ to $100\%$ solar) assuming a 5\,Myr constant-SFH. Different symbols/colors correspond to assumed gas electron temperatures of $10^{4}\,$K (blue), $1.5\times10^{4}\,$K (orange), and $2.5\times10^{4}\,$K (green).}
  \label{fig_fesc_vs_Z}
\end{figure}

The observational constraints for U37126 are overplotted in Fig.~\ref{fig_BPASS_predictions}. These favor 
stellar populations with very high $f_{\rm esc}$ along with young ages, $\simeq 2-10$\,Myr for constant star formation or $\simeq 1-3$\,Myr for single burst models. Thus, our results indicate that the lack of nebular emission in the MIRI/LRS spectrum is primarily driven by an exceptionally high $f_{\rm esc}$ and very young stellar populations, rather than by an evolved population with reduced LyC production. While single stellar population models (dashed red in Fig.~\ref{fig_BPASS_predictions}) with ages $>6$\,Myr can match the observed $EW_{0}(\mathrm{H\alpha})$ upper limit, they cannot reproduce simultaneously the observed steep UV slope and weak Balmer break.
As a simple estimate, we consider a 5\,Myr constant-SFH BPASS model redshifted to $z=10.255$ and scaled to match the observed spectrum. This scaled BPASS model, shown in Figure~\ref{fig_sed} (violet), has $\log (Q_{\rm H}/{\rm s^{-1}})\simeq 54.4$ (de-lensed) and a high ionizing photon production efficiency, log($\xi_{\rm ion} / \rm Hz\, erg^{-1}) \simeq 25.75$. For comparison, if we use the H$\alpha$ limit assuming $f_{\rm esc}=0$, we would obtain log($\xi_{\rm ion} / \rm Hz\, erg^{-1}) \leq 25.02$ (2$\sigma$), i.e., underpredicting its true $\xi_{\rm ion}$ by $>0.73$\,dex.
Together with our upper limit on the H$\alpha$ luminosity ($L\rm \,(H\alpha) \leq 4.5\times10^{41}$\,erg\,s$^{-1}$, de-lensed), our results indicate an $f_{\rm esc} \geq 86\%$ at 3$\sigma$ for U37126, assuming a stellar metallicity of $15\%\,Z_{\odot}$. 
Figure~\ref{fig_fesc_vs_Z} shows the corresponding 3$\sigma$ limits on the LyC escape fraction derived using the same methodology across a range of stellar metallicities ($0.5\%$ to $100\%\,Z_{\odot}$) and gas electron temperatures ($1.0$–$2.5 \times 10^{4}\,$K).
At low metallicities ($\leq 0.15\,Z_{\odot}$), we find $f_{\rm esc} \geq 86$–$91\%$ (3$\sigma$), whereas at higher metallicities ($\sim 0.5$–$1.0\,Z_{\odot}$), the limits decrease to $f_{\rm esc} \geq 69$–$80\%$ (3$\sigma$). We note, however, that other independent indicators (e.g., $\beta_{\rm UV}$) still favor high escape fractions at higher metallicities ($f_{\rm esc} \gtrsim 90\%$; see Figures~\ref{fig_appendix_A1} and \ref{fig_appendix_A2}).

In summary, our results point to a high LyC escape fraction largely independent of the assumed stellar population properties ($f_{\rm esc} \geq 86\%$ at 3$\sigma$ for our fiducial model). We emphasize that, given the limited observational constraints, these conclusions are also somewhat agnostic to the nature of the ionizing source. In particular, whether the ionizing radiation is dominated by a young stellar population or includes a contribution from an AGN, the combination of a hot ionizing source ($\beta_{\rm UV} \simeq -2.9$) and the lack of detectable nebular emission implies that only a small fraction of ionizing photons are absorbed by the interstellar medium.

\subsection{Spectral energy distribution and global properties}\label{SED}
    
We perform spectral energy distribution (SED) fitting using the SpectroPhotometric version of the CIGALE code \citep[V.2022.1;][]{Burgarella2005, Boquien2019, Burgarella2025A&A...699A.336B}, incorporating all NIRCam photometry, the NIRSpec/PRISM spectrum, as well as MIRI/LRS measurements of the [O\,{\sc iii}] $\lambda$5008 and H$\alpha$ flux upper limits and the continuum emission between $\lambda_{\rm obs}=4.8-5.6\,\mu$m, $5.85-6.73\,\mu$m, and $>7.0\,\mu$m (see Section~\ref{spectral_properties}). We have excluded from the fit all observations below $\lambda_{\rm rest} < 1400$\,\AA\ since they may be affected by IGM absorption and the effect is not properly handled by CIGALE. 

The star formation history (SFH) is modeled assuming constant star formation with ages varying from 1 to 20\,Myr in 1\,Myr steps. 
We adopt stellar population models from \cite{Bruzual&Charlot+03}, assuming a \cite{Chabrier+03} IMF and metallicities from $Z/Z_{\odot} = 2-20\%$. The ionization parameter ranges from log($U$) = $-3.0$ to $-1.5$ in 0.5\,dex steps. We adopt the Milky Way dust extinction law from \cite{Cardelli+89} with $R_{\rm V} = 3.1$ as the dust attenuation law for the nebular emission and the \cite{Calzetti+00} for the stellar emission. The color excess of the nebular gas is allowed to vary from 0 to 0.5\,mag. The escape of LyC photons is allowed to vary from 0 to 0.999. Finally, we add 10\% uncertainty to all dataset points to account for cross-calibration systematics between NIRSpec, NIRCam, and MIRI.

\begin{figure*}
  \centering
  \includegraphics[width=0.98\textwidth]{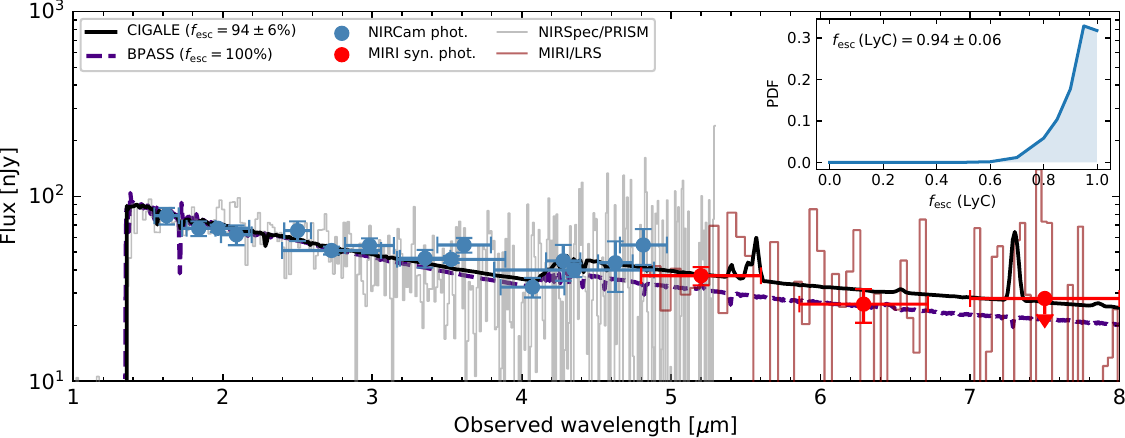}
  \caption{Best-fit spectral energy distribution (SED) of U37126 derived with CIGALE (black solid curve) and a pure stellar 5\,Myr-old BPASS model (CSFH with $Z/Z_{\odot}=15\%$, violet dashed line). The NIRSpec/PRISM and MIRI/LRS spectra are shown in grey and red, respectively, while NIRCam photometric measurements and MIRI/LRS synthetic photometry are indicated by blue and red circles. The inset panel displays the posterior probability density function of the Lyman-continuum escape fraction.}
  \label{fig_sed}
\end{figure*}

Our best-fit SED model ($\chi^{2}_{\nu} = 0.64$), shown in Figure~\ref{fig_sed}, is characterized by a stellar population with a constant star-formation rate $\rm SFR = 9.6 \pm 4.6$\,M$_{\odot}$\,yr$^{-1}$ (10\,Myr weighted), a stellar mass log($M_{\star}/\rm M_{\odot}) = 7.77 \pm 0.06$, and an age of $6.8\pm1.6$\,Myr (where global properties are corrected for magnification, assuming $\mu = 2.19$). This yields a specific star-formation rate $\rm sSFR = 160 \pm 80$\,Gyr$^{-1}$. The color excess is negligible, $E(B-V)=0.01\pm0.01$\,mag., as expected given the very steep UV slope of the best-fit model, $\beta_{\rm UV} = -2.85 \pm 0.05$ (and consistent with our measurements in Section ~\ref{spectral_properties}). The nebular metallicity is found to be $Z/Z_{\odot}=7.8 \pm 7.1 \%$.

The best-fit model also provides a very high LyC escape fraction, $f_{\rm esc} = 0.94 \pm 0.06$ (Figure~\ref{fig_sed}, inset panel), consistent with the results obtained in Section~\ref{predictions}. 
Finally, using the de-magnified size of $r_{\rm eff}=61\pm6$\,pc derived next in Section~\ref{morphology}, we obtain a stellar mass and SFR surface densities of $\log (\Sigma_{M\star} / M_{\odot}\,\rm pc^{-2}) = 3.40\pm0.10$ and $\log (\Sigma_{\rm SFR} / M_{\odot}\,\rm yr^{-1}\, kpc^{-2}) = 2.61\pm0.22$, respectively. These values are substantially higher than those of typical star-forming galaxies at high-$z$ \citep[e.g.,][]{Morishita2024ApJ...963....9M} and are instead comparable to those found in young massive star clusters and globular clusters \citep[e.g.,][]{Vanzella+23}, as well as in some of the most extreme sources at the highest redshifts \citep{Castellano2022ApJ...938L..15C,Tacchella+23,Naidu2025arXiv250511263N}. These properties are summarized in Table~\ref{table1}.

\subsection{Morphology and size}\label{morphology}

\begin{figure}
  \centering
  \includegraphics[width=0.46\textwidth]{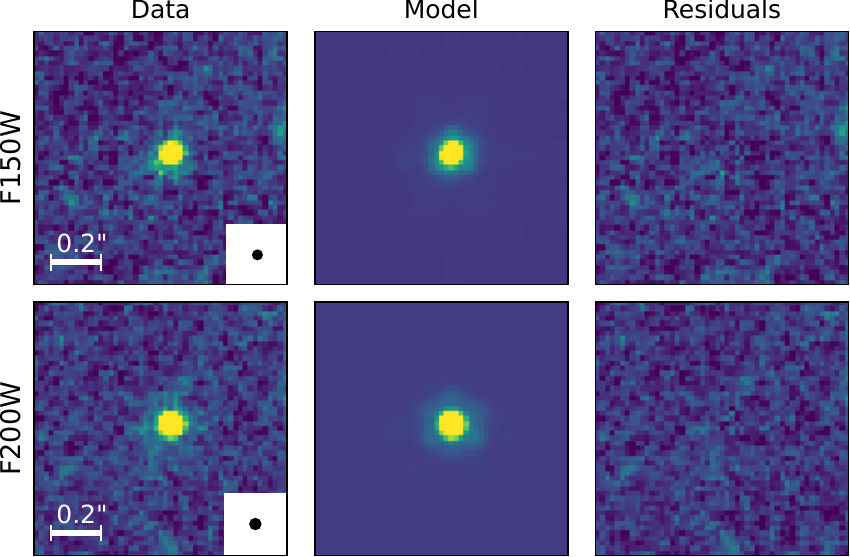}
  \caption{NIRCam F150W (top) and F200W (bottom) $1^{\prime\prime} \times 1^{\prime\prime}$ cutouts of U37126. Middle and right panels show the S\'ersic best-fit models obtained with \texttt{PySersic} and the corresponding residuals, respectively. The PSFs are shown in the left panels.}
  \label{fig_pysersic}
\end{figure}

U37126 shows a compact morphology in the NIRCam imaging, as illustrated in the top panels of Fig.~\ref{fig_pysersic}.
To quantify its structural properties, we model the light distribution using \texttt{PySersic} \citep{Pasha2023}.
\texttt{PySersic} performs forward modeling of galaxy morphologies with S\'ersic profiles convolved with a supplied point-spread function (PSF), and employs a Bayesian framework to explore the posterior distribution of all parameters and their degeneracies.

Empirical PSFs from the UNCOVER DR4 release \citep{Suess2024ApJ...976..101S, Weaver2024ApJS..270....7W} are available for all NIRCam bands, but only at a pixel scale of $0.04^{\prime\prime},\mathrm{pix}^{-1}$.
Because the short-wavelength NIRCam images (e.g., F150W and F200W) have been resampled in UNCOVER DR4 to a pixel scale of $0.02^{\prime\prime}$\,pix$^{-1}$, we generate PSFs for these bands using \texttt{STPSF}\footnote{\url{https://stpsf.readthedocs.io/en/latest/}} \citep{Perrin+14}. We follow \cite{Morishita2024ApJ...963....9M} and \cite{Weibel2024MNRAS.533.1808W} and set the
\texttt{jitter\_sigma} parameter in \texttt{WebbPSF} to 0.022, which has been shown to best reproduce the observed NIRCam PSFs.

We fit 2D S\'ersic models, allowing the S\'ersic index to vary between $0.5$ and $6.0$, while leaving the total flux, effective radius, ellipticity, and position angle free. Given the significantly higher SNR in the NIRCam broadband filters relative to the medium bands, we restrict our morphological analysis to the broadband imaging only, also excluding F444W image for the same reason (low SNR).
Each fit is performed on a $1^{\prime\prime}\times1^{\prime\prime}$ cutout, all centered on U37126.

For the short-wavelength NIRCam images, we measure effective radii of $r_{\rm eff} = 24\pm2$\,mas in F150W and $r_{\rm eff} = 22\pm2$\,mas in F200W,
respectively (Figure~\ref{fig_pysersic}), and a S\'ersic index of $n=1.64\pm0.26$. We adopt the F200W value, as this band probes the rest-frame UV at $\simeq 1750$\AA. Given that U37126 is moderately magnified by A2744, with $\mu = 2.19 \pm 0.05$ \citep{Furtak2023MNRAS.523.4568F, Price2025ApJ...982...51P}, we obtain a de-magnified radius of $r_{\rm eff} = 61\pm6$\,pc at $z=10.255$. 
At longer wavelengths, the best-fit models yield $r_{\rm eff} \lesssim 0.5$\,pix (or $r_{\rm eff} \lesssim 0.02^{\prime \prime}$). Thus, U37126 appears unresolved in the rest-optical with $r_{\rm eff} < 160$\,pc (or $<100$\,pc after lensing correction ).

\section{Discussion}\label{Sect4:disc}

\subsection{Comparison with other strong confirmed LyC emitters and $z>6$ candidates}

\begin{figure}
  \centering
  \includegraphics[width=0.49\textwidth]{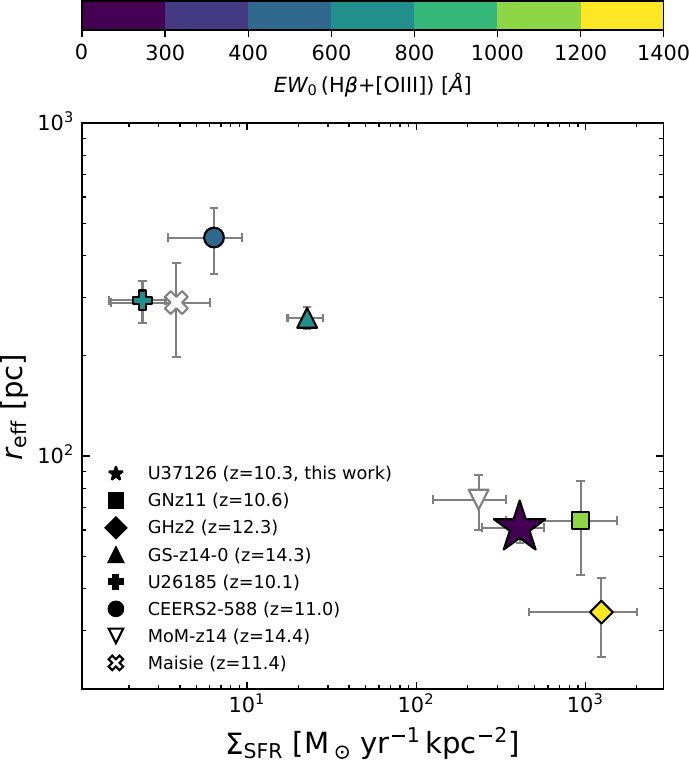}
  \caption{Effective radius ($r_{\rm eff}$) as a function of star-formation rate surface density ($\Sigma_{\rm SFR}$) for galaxies at $z>10$ with MIRI constraints on $EW_{0}$\,(H$\beta$+[O\,{\sc iii}]) (color-codded, from: \citealt{Castellano2022ApJ...938L..15C, Goulding2023ApJ...955L..24G,  Tacchella+23, Calabro2024,  Carniani+24, Alvarez2025A&A...695A.250A, Helton2025arXiv251219695H, Zavala2025NatAs...9..155Z, Harikane2026arXiv260121833H}, and \citealt{alvarez-marquez2026arXiv260202323A}). Symbols denote individual sources while the open symbols indicate sources without available $EW_{0}$ measurements \citep{Arrabal2023Natur.622..707A, Naidu2025arXiv250511263N}. U37126 (star) occupies the compact, high-$\Sigma_{\rm SFR}$ regime but shows unusually weak nebular emission with $EW_{0}$\,(H$\beta$+[O\,{\sc iii}]$)<296$\,\AA~(3\,$\sigma$) compared to other compact systems (e.g., GNz11 and GHz2), consistent with a high LyC escape fraction.
 }
  \label{fig_highz}
\end{figure}

Our results strongly support a scenario in which U37126 both produces and leaks large amounts of LyC photons. These constraints are driven by the combination of its extremely steep UV continuum and the non-detection of nebular emission, and indirectly, from the very sharp Ly$\alpha$ break. Although extreme, similar properties have been identified with JWST in a small number of sources at $z\sim 6-10$ \citep{Topping2022ApJ...941..153T, Hainline2024ApJ...976..160H, Giovinazzo2025arXiv250701096G, Yanagisawa2025ApJ...988...86Y, Jecmen2026arXiv260119995J}. These galaxies share comparably blue UV continua and unusually weak nebular emission, pointing toward very young stellar populations with very high LyC $f_{\rm esc}$. Interestingly, several of these also show high SFRs within extremely compact morphologies ($r_{\rm eff}<260$\,pc). This results in very high $\Sigma_{\rm SFR}$, comparable to that inferred for U37126, $\log(\Sigma_{\rm SFR}/{\rm M_\odot\,yr^{-1}\,kpc^{-2}})\simeq2.6$, and consistent with expectations for strong LyC leakage \citep{Sharma2017MNRAS.468.2176S, Naidu2020ApJ...892..109N}.

Recent JWST studies have further suggested a dichotomy among galaxies at $z \gtrsim 10$ \citep{Harikane2025ApJ...980..138H, Roberts-Borsani2025arXiv250821708R}, separating extended systems with relatively weak emission lines, in some cases accompanied by Balmer breaks \citep{Harikane2026arXiv260121833H, alvarez-marquez2026arXiv260202323A}, that may be undergoing a recent decline in star formation \citep[e.g.,][]{Helton2025arXiv251219695H}, from very compact galaxies ($r_{\rm eff} \lesssim 100$\,pc) experiencing recent bursts and exhibiting strong UV and/or optical emission lines \citep[e.g.,][]{Bunker+23, Castellano2024, Alvarez2025A&A...695A.250A, Zavala2025NatAs...9..155Z}. 
Figure~\ref{fig_highz} shows the effective radius and star-formation rate surface density of the small sample of $z>10$ sources with MIRI constraints on rest-optical emission, with the color scale encoding the rest-frame equivalent width $EW_{0}$(H$\beta$+[O\,{\sc iii}]). Compact systems generally occupy the regime of high $\Sigma_{\rm SFR}$ and large equivalent widths. U37126 lies in this compact, high-$\Sigma_{\rm SFR}$ regime, yet it shows unusually weak nebular emission, which is naturally explained by its very high LyC escape fraction.

To further place U37126 in context, we compare its observed properties with those of confirmed LyC emitters at lower redshifts ($z<4$), where LyC escape can be measured directly. We first note that the majority of known LyC emitters exhibit relatively modest escape fractions and UV slopes $\beta_{\rm UV}>-2.5$ \citep[e.g.,][]{Izotov2018MNRAS.478.4851I, Steidel2018ApJ...869..123S, Marques-Chaves2021MNRAS.507..524M, Flury2022ApJ...930..126F, Kerutt2024A&A...684A..42K}. Only a small subset shows $f_{\rm esc} > 50\%$ and nearly all of these exhibit strong nebular emission, with typical H$\beta$ equivalent widths of the order of $\sim100-300$\,\AA~\citep[][]{deBarros2016A&A...585A..51D, Izotov2016MNRAS.461.3683I, Vanzella2016ApJ...825...41V, Izotov2018MNRAS.478.4851I, Rivera-Thorsen2019Sci...366..738R, Flury2022ApJ...930..126F}. An exception is the strong LyC leaker J1316+2614 at $z=3.6$, which, despite being much brighter, has a measured $f_{\rm esc}\simeq 87\%$, $EW_{0}\,({\rm H}\beta) \simeq 35$\,\AA, and $\beta_{\rm UV} \simeq -2.6$ \citep{Marques-Chaves2022MNRAS.517.2972M}, representing the closest known analog to U37126 and to other $z>6$ strong LyC emitting candidates with steep UV slopes and weak nebular emission.

At face value, the observed high $f_{\rm esc}$ and strong nebular emission in some confirmed strong LyC emitters appear at odds with the predictions and with the properties of U37126 (Fig.~\ref{fig_BPASS_predictions}). However, these observations are not necessarily contradictory. The measured $f_{\rm esc}$ in low-$z$ LyC emitters probes LyC escape along the line of sight toward the young stars, whereas nebular emission traces a more global, $4\pi$-averaged escape fraction. Strongly anisotropic LyC leakage can therefore result in high apparent $f_{\rm esc}$ values while still producing prominent nebular emission \citep[e.g.,][]{Flury2022ApJ...930..126F, Jaskot2024ApJ...972...92J}. In addition, LyC flux measurements close to the Lyman limit may be boosted by nebular bound-free emission \citep{Inoue2010MNRAS.401.1325I, Simmonds2024MNRAS.530.2133S}, potentially leading to an overestimation of the stellar $f_{\rm esc}$. As recently shown by \cite{Izotov2025A&A...704A..19I}, this effect appears to be significant in J1243+4646, the strongest LyC emitter at $z\simeq0.3$ and also a strong line emitter ($f_{\rm esc} \simeq 72\%$ and $EW_{0} \rm (H\beta)\simeq 221$\,\AA, \citealt{Izotov2018MNRAS.478.4851I}).

These results thus suggest that sources like U37126, characterized by very steep UV slopes with faint nebular emission, may be rare or absent at $z<4$. However, it remains unclear whether this reflects a genuinely low number density at lower redshifts, possibly due to the specific physical conditions required to form such systems (see Section~\ref{section_conditions}), or whether they have been systematically overlooked in LyC surveys due to selection effects \citep[e.g.,][]{Bergvall2013A&A...554A..38B}, as the confirmation of spectroscopic redshifts generally relies on strong nebular emission.

\subsection{Conditions for forming ``ISM-naked'' starbursts with nearly unity LyC escape fraction}
\label{section_conditions}

The nature of sources like U37126 is intriguing, as typical starburst galaxies are gas-rich and exhibit strong nebular emission. The stringent $3\sigma$ upper limits on the equivalent widths, together with the very steep UV continuum and a weak/flat Balmer break strength, indicate that nebular emission is intrinsically weak and at most a small fraction of LyC photons is being reprocessed by surrounding gas.

To obtain a rough estimate of the characteristic size of the ionized region associated with U37126, we compute the Str\"omgren radius,
\[
R_{\rm S} = \left( \frac{3\,Q_{\rm H}}{4\pi\,\alpha_{\rm B}\,n_{\rm H}^2} \right)^{1/3},
\]
where $Q_{\rm H} \approx 10^{54.4}$\,s$^{-1}$ (Section~\ref{Sect:results_dis}), $n_{\rm H}$ is the hydrogen number density of the surrounding gas, and $\alpha_{\rm B}$ is the case-B recombination coefficient. For typical electron temperatures $T_{\rm e} \simeq (1$--$2)\times10^{4}$\,K and gas densities $n_{\rm H} \simeq 10^{2}$--$10^{3}\,\mathrm{cm^{-3}}$, we infer Str\"omgren radii of $R_{\rm S} \approx 50-200$\,pc. These scales are comparable to, or exceed, the size of the stellar emission ($r_{\rm eff} \simeq 61$\,pc; Section~\ref{morphology}). Under such extreme conditions, any substantial gas reservoir located within or around U37126 would be expected to be efficiently ionized and, therefore, most likely detectable in the NIRSpec and MIRI spectra. 

One possible exception is a scenario in which the ionized gas is extremely diffuse, since line luminosities scale as $L \propto n_{\rm e}^{2}$. Even if all LyC photons were fully reprocessed within H\,{\sc ii} regions, the non-detection of H$\alpha$ emission could, in principle, be explained by an extremely low electron density ($n_{\rm e} \lesssim 10^{-2}$\,cm$^{-3}$). However, such conditions are rarely observed and are particularly unlikely for U37126 ($z = 10.255$), given the expected increase of electron density toward higher redshifts \citep[e.g.,][]{Isobe+23}. Moreover, the extreme stellar mass and SFR surface densities of U37126 would instead favor high electron densities \citep[e.g.,][]{Reddy2023ApJ...951...56R, Reddy2023ApJ...952..167R}. Consistent with this expectation, nearly all sources at $z > 10$ with sizes comparable to that of U37126 ($r_{\rm eff} \approx 61$\,pc) exhibit evidence for extremely high gas densities \citep{Harikane2025ApJ...980..138H}, several of them shown in the bottom right corner of Figure~\ref{fig_highz} (e.g., GNz11, GHz2).
Taken together, the absence of detectable nebular emission in U37126 suggests that the galaxy is largely depleted of its interstellar medium. U37126 thus behaves as an ISM-free starburst, raising the question of how such conditions can be achieved.

One possibility is that, during its intense star-formation episode, the bulk of the natal gas in U37126 was rapidly and efficiently converted into stars, leaving little residual gas available to absorb and reprocess LyC photons. \cite{Dekel2023} (see also \citealt{Li+Dekel2023}) predict that extremely high gas surface densities ($\Sigma_{\rm gas} \gtrsim 3\times10^{3}\,M_{\odot}\,\mathrm{pc^{-2}}$) can lead to the formation of ``feedback-free starbursts'' (FFB), in which gas clouds collapse on very short free-fall timescales ($\sim1$\,Myr). This makes star formation very efficient, since the cloud collapse occurs before the onset of mechanical feedback (e.g., SNe). While the gas surface density of U37126 cannot be measured directly, the observed stellar mass surface density, $\log(\Sigma_{M\star}/M_{\odot}\,\mathrm{pc^{-2}})=3.40\pm0.10$, suggests a pre-star-formation gas surface density of at least comparable magnitude, thus satisfying the condition for an FFB. Observationally, high star-formation efficiencies have been inferred ($>40\%$; \citealt{Dessauges-Zavadsky2025A&A...693A..17D}) and predicted ($>70\%$; \citealt{Marques-Chaves+24_SFE}) to explain the strong LyC escape directly measured in the strong leaker J1316+2614 at $z=3.6$ with $f_{\rm esc}\approx 87\%$ and $EW_{0}\,({\rm H}\beta) \simeq 35$\,\AA~\citep{Marques-Chaves2022MNRAS.517.2972M}. 

An alternative, though not mutually exclusive, scenario is that strong feedback has removed gas and dust from the stellar core of U37126, and potentially to larger distances.\footnote{Given the widths of the NIRSpec/MSA and MIRI/LRS apertures of $\simeq 0.2^{\prime\prime}$ and $0.5^{\prime\prime}$, respectively, within which no nebular emission appears detected. This corresponds to $\sim0.8$\,kpc and $\sim2.1$\,kpc at $z=10.25$, i.e., much larger than the size of U37126 ($r_{\rm eff}\simeq 61$\,pc).} In this context, \cite{Ferrara23} proposed that radiation pressure on dust and gas can drive powerful radiative feedback during super-Eddington phases. Such conditions are expected in the early stages of a starburst, when the sSFR exceeds a critical threshold of $\gtrsim 25$\,Gyr$^{-1}$ \citep{Fiore23}, which is satisfied in this source ($\simeq 160$\,Gyr$^{-1}$). These phases may also facilitate substantial LyC leakage, as shown recently by \cite{Ferrara2025OJAp....8E.125F}. Furthermore, recent hydrodynamic simulations have shown that strong radiative outflows can only be efficiently launched in very dense systems with high star-formation efficiencies, leading simultaneously to strong LyC leakage \citep{Menon2025ApJ...987...12M}. 
Observational evidence for such extreme outflows has been reported in a handful of powerful starbursts with elevated sSFR \citep{Crespo2025arXiv251114658C, Marques-Chaves2025arXiv251012411M}, including in confirmed low-$z$ LyC emitters \citep[e.g.,][]{Komarova2025ApJ...994..192K}. However, these systems still exhibit intense nebular emission, unlike U37126. 

In summary, the absence of detectable gas in U37126 may reflect an evolutionary sequence in which exceptionally efficient star formation rapidly consumes the natal gas, followed by intense feedback that expels the remaining material, including dust, from the stellar core. In the absence of subsequent gas accretion, U37126 may then evolve rapidly into a post-starburst or recently quenched system, like the ones recently identified by JWST at high-$z$ \citep[e.g.,][]{Looser2024Natur.629...53L}.

\subsection{Implications for cosmic reionization}

The discovery of powerful ionizing sources such as U37126 has important implications for cosmic reionization. The ionizing photon budget is commonly expressed through the comoving ionizing emissivity, $\dot{n}_{\rm ion}$, defined as $\dot{n}_{\rm ion} = \rho_{\rm UV}\,\xi_{\rm ion}\,f_{\rm esc}$, where $\rho_{\rm UV}$ is the integral of the UV luminosity function \citep{Robertson+13}. Observations indicate that the average galaxy population in reasonably complete samples down to $M_{\rm UV} \approx -17$ at $z \gtrsim 6$ exhibits $\log(\xi_{\rm ion}/\mathrm{Hz\,erg^{-1}}) \simeq 25.2 - 25.3$ \citep[e.g.,][]{Choustikov2024MNRAS.529.3751C, Mascia2024A&A...685A...3M, Simmonds2024, Pahl2025ApJ...981..134P, Begley2026MNRAS.545S1995B}, consistent with, or only marginally higher than, pre-JWST canonical values \citep[e.g.,][]{Robertson2015ApJ...802L..19R}. For these $\xi_{\rm ion}$, reionization models typically require population-averaged escape fractions of $f_{\rm esc} \simeq 10-20$\% \citep{Robertson2015ApJ...802L..19R}, broadly supported by inferences using indirect LyC indicators \citep{Mascia+23, Jaskot2024ApJ...973..111J, Jecmen2026arXiv260119995J} but with large scatter and sensitive to assumptions about the faint-end slope and cutoff of the UV luminosity function \citep[e.g.,][]{Finkelstein2019ApJ...879...36F, Korber2026arXiv260119989K}. For simplicity, we adopt a fiducial scenario in which reionization is sustained by an averaged galaxy population characterized by $f_{\rm esc} =15\%$ and $\log(\xi_{\rm ion}/\mathrm{Hz\,erg^{-1}}) = 25.25$. 

\begin{figure}
  \centering
  \includegraphics[width=0.48\textwidth]{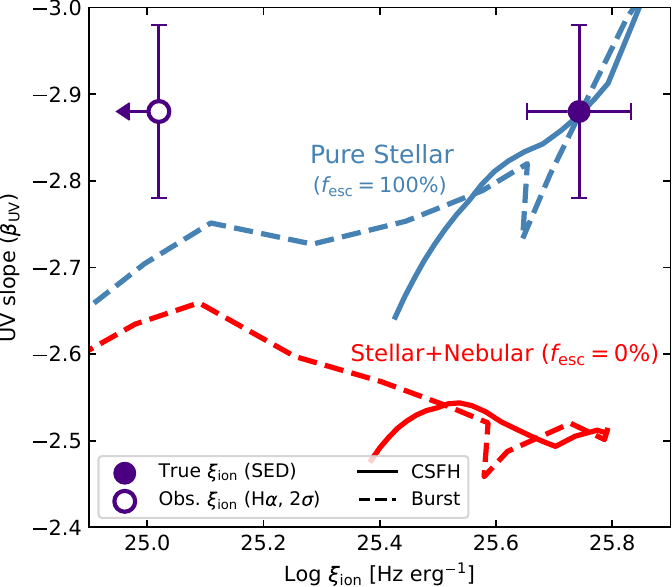}
  \caption{Relationship between the UV slope ($\beta_{\rm UV}$) and the ionizing photon production efficiency ($\xi_{\rm ion}$) for stellar population synthesis models with (red) and without (blue) the contribution of nebular emission, i.e., representing the extreme cases of $f_{\rm esc}=0$ and $f_{\rm esc}=1$, respectively. 
  Filled and open circles indicate $\xi_{\rm ion}$ derived from SED fitting and from the H$\alpha$ flux limit assuming $f_{\rm esc}=0$, respectively.  }
  \label{fig_xion}
\end{figure}

We now consider a simple, illustrative scenario in which the ionizing budget is fully dominated by sources similar to U37126. Owing to their exceptionally high $\xi_{\rm ion}$ and near-unity $f_{\rm esc}$, such systems could contribute disproportionately to $\dot{n}_{\rm ion}$ despite being rare. 
Requiring the ionizing emissivity contributed by these extreme sources to match that of the fiducial galaxy population yields:
\[
\left(\frac{f_{\rm esc}}{15\%}\right)
\left(\frac{\xi_{\rm ion}}{10^{25.25}}\right)
=
f_{N}
\left(\frac{f_{\rm esc}}{\geq 86\%}\right)
\left(\frac{\xi_{\rm ion}}{10^{25.75}}\right),
\]
where $f_{N}$ is the fraction of such extreme sources relative to the total galaxy population. Adopting the properties inferred for U37126, $\log(\xi_{\rm ion}/\mathrm{Hz\,erg^{-1}}) \simeq 25.75$ and $f_{\rm esc} \geq 86\%$ (3$\sigma$; Section~\ref{fig_BPASS_predictions}), we find $f_{N} \lesssim 6\%$.

While this scenario is intentionally simplistic, implicitly assuming that such extreme values of $\xi_{\rm ion}$ and $f_{\rm esc}$ are independent of UV luminosity and populate the full luminosity function, it nevertheless illustrates that even a tiny fraction of powerful ionizing sources like U37126 could contribute significantly, or even dominate the ionizing photon budget during reionization. This picture is consistent with the results of \cite{Papovich2025arXiv250508870P}, who find that the majority ($\simeq 82\%$) of galaxies at $z\simeq5-8$ exhibit negligible LyC escape ($f_{\rm esc} < 1\%$), while only a small subset are strong LyC emitters, suggesting a highly bimodal distribution in $f_{\rm esc}$ where reionization may be driven by rare but efficient leakers.

In this regard, recent JWST observations are revealing a progressive steepening of UV slopes with increasing redshift \citep[e.g.,][]{Topping2022ApJ...941..153T, Cullen2024MNRAS.531..997C, Austin2025ApJ...995...43A, Dottorini2025A&A...698A.234D}, with several sources at $z\gtrsim6$ exhibiting extremely blue continua ($\beta_{\rm UV}<-2.8$). As illustrated in Fig.~\ref{fig_xion}, such UV slopes require very young stellar populations, implying elevated ionizing photon production efficiencies, $\xi_{\rm ion}\gtrsim10^{25.6}$\,Hz\,erg$^{-1}$, together with high LyC escape fractions ($f_{\rm esc}\gtrsim50\%$; see Fig.~\ref{fig_BPASS_predictions}). Moreover, the statistical analysis of \cite{Topping+24} identified a small but non-negligible fraction ($\simeq3.4\%$) of high-redshift sources with $\beta_{\rm UV}<-2.8$ and indications of weak nebular emission, closely resembling U37126, which is comparable to the $f_{N} \lesssim6\%$ of extreme LyC emitters required in our illustrative scenario. 
If spectroscopically confirmed, such systems could already account for a substantial fraction ($\gtrsim50\%$) of the total ionizing photon budget during cosmic reionization.

\section{Summary and Conclusions}\label{Sect:conclusion_Summary}

In this work, we have presented very deep ($\simeq 11$\,h on-source) JWST/MIRI LRS rest-frame optical spectroscopy of U37126, a $M_{\rm UV} = -20.10$, mildly lensed ($\mu\sim 2.2$) galaxy at $z = 10.255$ previously identified with NIRSpec spectroscopy by \cite{Fujimoto2024ApJ...977..250F}. 

The source exhibits an exceptionally steep UV continuum slope, $\beta_{\rm UV} \simeq -2.9$, a weak Balmer break, a sharp Lyman-$\alpha$ break, and intrinsically faint nebular emission. Despite the clear detection of the continuum in both the NIRSpec/PRISM and MIRI/LRS spectra, no recombination or metal emission lines are detected. We derive stringent $3\sigma$ rest-frame equivalent-width upper limits of $\leq 64$\,\AA, $\leq 174$\,\AA, and $\leq 400$\,\AA\ for H$\beta$, [O\,{\sc iii}]\,$\lambda5008$, and H$\alpha$, respectively.

Combining these observational constraints with synthetic stellar population models, we have shown that the spectral properties of U37126 require both extremely young stellar ages and a very high Lyman-continuum escape fraction. Our results indicate ages $\leq 2$\,Myr for instantaneous bursts or $\leq 10$\,Myr for constant star formation, implying a very high ionizing photon production efficiency ($\log(\xi_{\rm ion}/\mathrm{Hz\,erg^{-1}}) \geq 25.6$). From the H$\alpha$ luminosity limit, we derive a conservative lower limit of $f_{\rm esc} \geq 86\%$ (3$\sigma$), while independent SED fitting favors $f_{\rm esc} = 0.94 \pm 0.06$. 

U37126 is extremely compact, with a de-lensed effective radius of $r_{\rm eff} \simeq 61$\,pc. The best-fit SED yields a (de-lensed) stellar mass of $M_{\star} \simeq 10^{7.8}\,M_{\odot}$ and a star-formation rate of $\mathrm{SFR} \simeq 10\,M_{\odot}\,\mathrm{yr^{-1}}$. This yields very high stellar mass and star-formation-rate surface densities of $\log(\Sigma_{M\star}/M_{\odot}\,\mathrm{pc^{-2}}) \simeq 3.4$ and $\log(\Sigma_{\mathrm{SFR}}/M_{\odot}\,\mathrm{yr^{-1}\,kpc^{-2}}) \simeq 2.6$, respectively, which are only comparable to those found in young massive star clusters and the most extreme sources at the highest redshifts. 

Together with the lack of detectable nebular emission, these properties suggest that U37126 is undergoing an ISM-free starburst phase, in which the ISM is strongly depleted, and only a small fraction of ionizing photons are reprocessed by surrounding gas. Such conditions may result from both an extremely efficient gas-to-star conversion and/or strong feedback that has efficiently cleared the ISM from its stellar core. While our results indicate that LyC photons escape efficiently from the ISM of U37126, recent constraints on the neutral hydrogen column density from the Ly$\alpha$ break ($\log(N_{\rm HI}/{\rm cm}^{-2}) = 19.2 \pm 1.5$) indicate that substantial neutral gas may still be present along the line of sight. Therefore, while our observations favor a very high escape fraction from the galaxy itself, it remains unclear whether the escaping LyC radiation has already reached and ionized its surrounding environment on larger scales.

Although systems like U37126 are likely rare, their extreme properties, combining both a high production and escape of LyC photons, suggest that they could contribute disproportionately to the ionizing photon budget during cosmic reionization. Even a small fraction of such sources ($\sim 3\%$--$6\%$)  could account for a substantial share ($\sim 50\%$--$100\%$) of the required ionizing emissivity. This scenario is consistent with emerging JWST evidence for increasingly blue UV continua and weak nebular emission among the highest-redshift galaxies. If confirmed in larger statistical samples, these compact, highly efficient LyC emitters like U37126 may represent a key population responsible for driving and sustaining cosmic reionization.

\begin{acknowledgements}

We thank the anonymous referee for the useful comments and suggestions. This work is based on observations made with the NASA/ESA/CSA James Webb Space Telescope. The data were obtained from the Mikulski Archive for Space Telescopes at the Space Telescope Science Institute, which is operated by the Association of Universities for Research in Astronomy, Inc., under NASA contract NAS 5-03127 for JWST. These observations are associated with program \#8051.

J.A.-M., C.P.-J., B.R.P. acknowledge support from grant PID2024-158856NA-I00, J.A.-M., L.C., C.P.-J., B.R.P. acknowledge support from grant PID2021-127718NB-100, P.G.P.-G. acknowledges support from grant PID2022-139567NB-I00 from the Spanish Ministry of Science and Innovation/State Agency of Research MCIN/AEI/10.13039/501100011033 and by “ERDF A way of making Europe”. J.A.-M., L.C., C.P.-J., B.R.P., P.G.P.-G. acknowledge support by grant CSIC/BILATERALES2025/BIJSP25022. M.C. acknowledges INAF GO Grant 2024 "Revealing the nature of bright galaxies at cosmic dawn with deep JWST spectroscopy". T.H. was supported by JSPS KAKENHI 25K00020. Y.H. acknowledges support from the Japan Society for the Promotion of Science (JSPS) Grant-in-Aid for Scientific Research (24H00245), the JSPS Core-to-Core Program (JPJSCCA20210003), and the JSPS International Leading Research (22K21349). Y.F. is supported by JSPS KAKENHI Grant Numbers JP22K21349 and JP23K13149. D.L. was supported by research grants (VIL16599,VIL54489) from VILLUM FONDEN. P.S. acknowledges support from INAF RF2024 Large Grant "UNDUST: UNveiling the Dawn of the Universe with JWST"

The data were obtained from the Mikulski Archive for Space Telescopes at the Space Telescope Science Institute, which is operated by the Association of Universities for Research in Astronomy, Inc., under NASA contract NAS 5-03127 for \textit{JWST}; and from the \href{https://jwst.esac.esa.int/archive/}{European \textit{JWST} archive (e\textit{JWST})} operated by the ESDC.

This research made use of Photutils, an Astropy package for detection and photometry of astronomical sources \citep{larry_bradley_2022_6825092}.

\end{acknowledgements}

\bibliographystyle{aa} % style aa.bst
\bibliography{bibliography.bib} % your references Yourfile.bib

@ARTICLE{Choustikov2024MNRAS.529.3751C,
       author = {{Choustikov}, Nicholas and {Katz}, Harley and {Saxena}, Aayush and {Cameron}, Alex J. and {Devriendt}, Julien and {Slyz}, Adrianne and {Rosdahl}, Joki and {Blaizot}, Jeremy and {Michel-Dansac}, Leo},
        title = "{The Physics of Indirect Estimators of Lyman Continuum Escape and their Application to High-Redshift JWST Galaxies}",
      journal = {\mnras},
         year = 2024,
        month = apr,
       volume = {529},
       number = {4},
        pages = {3751-3767},
          doi = {10.1093/mnras/stae776},
archivePrefix = {arXiv},
       eprint = {2304.08526},
 primaryClass = {astro-ph.GA},
       adsurl = {https://ui.adsabs.harvard.edu/abs/2024MNRAS.529.3751C}
}

@ARTICLE{Mason2026A&A...705A.114M,
       author = {{Mason}, Charlotte A. and {Chen}, Zuyi and {Stark}, Daniel P. and {Yi Lu}, Ting and {Topping}, Michael and {Tang}, Mengtao},
        title = "{Constraints on the z {\ensuremath{\sim}} 6{\ensuremath{-}}13 intergalactic medium from JWST spectroscopy of Lyman-alpha damping wings in galaxies}",
      journal = {\aap},
         year = 2026,
        month = jan,
       volume = {705},
          eid = {A114},
        pages = {A114},
          doi = {10.1051/0004-6361/202553820},
archivePrefix = {arXiv},
       eprint = {2501.11702},
 primaryClass = {astro-ph.GA},
       adsurl = {https://ui.adsabs.harvard.edu/abs/2026A&A...705A.114M}
}

@ARTICLE{alvarez-marquez2026arXiv260202323A,
       author = {{{\'A}lvarez-M{\'a}rquez}, J. and {Colina}, L. and {Crespo-Gomez}, A. and {Kendrew}, S. and {Zavala}, J. and {Marques-Chaves}, R. and {Prieto-Jim{\'e}nez}, C. and {Abdurro'uf} and {Blanco-Prieto}, C. and {Boogaard}, L.~A. and {Castellano}, M. and {Fontana}, A. and {Fudamoto}, Y. and {Fujimoto}, S. and {Garc{\'\i}a-Mar{\'\i}n}, M. and {Harikane}, Y. and {Harish}, S. and {Hashimoto}, T. and {Hsiao}, T. and {Iani}, E. and {Inoue}, A.~K. and {Langeroodi}, D. and {Lin}, R. and {Melinder}, J. and {Napolitano}, L. and {Ostlin}, G. and {P{\'e}rez-Gonz{\'a}lez}, P.~G. and {Rinaldi}, P. and {Rodr{\'\i}guez Del Pino}, B. and {Santini}, P. and {Sugahara}, Y. and {Treu}, T. and {Varo-O'ferral}, A. and {Wright}, G.},
        title = "{PRISMS. UNCOVER-26185, a metal-poor SFG at z=10.05 with no evidence for a X-ray-luminous AGN}",
      journal = {arXiv e-prints},
         year = 2026,
        month = feb,
          eid = {arXiv:2602.02323},
        pages = {arXiv:2602.02323},
          doi = {10.48550/arXiv.2602.02323},
archivePrefix = {arXiv},
       eprint = {2602.02323},
 primaryClass = {astro-ph.GA},
       adsurl = {https://ui.adsabs.harvard.edu/abs/2026arXiv260202323A}
}

@ARTICLE{Papovich2025arXiv250508870P,
       author = {{Papovich}, Casey and {Cole}, Justin W. and {Hu}, Weida and {Finkelstein}, Steven L. and {Shen}, Lu and {Arrabal Haro}, Pablo and {Amor{\'\i}n}, Ricardo O. and {Backhaus}, Bren E. and {Bagley}, Micaela B. and {Bhatawdekar}, Rachana and {Calabr{\`o}}, Antonello and {Carnall}, Adam C. and {Cleri}, Nikko J. and {Daddi}, Emanuele and {Dickinson}, Mark and {Grogin}, Norman A. and {Holwerda}, Benne W. and {Jaskot}, Anne E. and {Koekemoer}, Anton M. and {Llerena}, Mario and {Lucas}, Ray A. and {Mascia}, Sara and {Pacucci}, Fabio and {Pentericci}, Laura and {P{\'e}rez-Gonz{\'a}lez}, Pablo G. and {Pirzkal}, Nor and {Raghunathan}, Srinivasan and {Seill{\'e}}, Lise-Marie and {Somerville}, Rachel S. and {Yung}, L.~Y. Aaron},
        title = "{Galaxies in the Epoch of Reionization Are All Bark and No Bite{\textemdash}Plenty of Ionizing Photons, Low Escape Fractions}",
      journal = {\apj},
         year = 2026,
        month = mar,
       volume = {1000},
       number = {1},
          eid = {111},
        pages = {111},
          doi = {10.3847/1538-4357/ae3b25},
archivePrefix = {arXiv},
       eprint = {2505.08870},
 primaryClass = {astro-ph.GA},
       adsurl = {https://ui.adsabs.harvard.edu/abs/2026ApJ..1000..111P}
}

@ARTICLE{Menon2025ApJ...987...12M,
       author = {{Menon}, Shyam H. and {Burkhart}, Blakesley and {Somerville}, Rachel S. and {Thompson}, Todd A. and {Sternberg}, Amiel},
        title = "{Bursts of Star Formation and Radiation-driven Outflows Produce Efficient LyC Leakage from Dense Compact Star Clusters}",
      journal = {\apj},
         year = 2025,
        month = jul,
       volume = {987},
       number = {1},
          eid = {12},
        pages = {12},
          doi = {10.3847/1538-4357/add2f9},
archivePrefix = {arXiv},
       eprint = {2408.14591},
 primaryClass = {astro-ph.GA},
       adsurl = {https://ui.adsabs.harvard.edu/abs/2025ApJ...987...12M}
}

@ARTICLE{Arrabal2023Natur.622..707A,
       author = {{Arrabal Haro}, Pablo and {Dickinson}, Mark and {Finkelstein}, Steven L. and {Kartaltepe}, Jeyhan S. and {Donnan}, Callum T. and {Burgarella}, Denis and {Carnall}, Adam C. and {Cullen}, Fergus and {Dunlop}, James S. and {Fern{\'a}ndez}, Vital and {Fujimoto}, Seiji and {Jung}, Intae and {Krips}, Melanie and {Larson}, Rebecca L. and {Papovich}, Casey and {P{\'e}rez-Gonz{\'a}lez}, Pablo G. and {Amor{\'\i}n}, Ricardo O. and {Bagley}, Micaela B. and {Buat}, V{\'e}ronique and {Casey}, Caitlin M. and {Chworowsky}, Katherine and {Cohen}, Seth H. and {Ferguson}, Henry C. and {Giavalisco}, Mauro and {Huertas-Company}, Marc and {Hutchison}, Taylor A. and {Kocevski}, Dale D. and {Koekemoer}, Anton M. and {Lucas}, Ray A. and {McLeod}, Derek J. and {McLure}, Ross J. and {Pirzkal}, Norbert and {Seill{\'e}}, Lise-Marie and {Trump}, Jonathan R. and {Weiner}, Benjamin J. and {Wilkins}, Stephen M. and {Zavala}, Jorge A.},
        title = "{Confirmation and refutation of very luminous galaxies in the early Universe}",
      journal = {\nat},
         year = 2023,
        month = oct,
       volume = {622},
       number = {7984},
        pages = {707-711},
          doi = {10.1038/s41586-023-06521-7},
archivePrefix = {arXiv},
       eprint = {2303.15431},
 primaryClass = {astro-ph.GA},
       adsurl = {https://ui.adsabs.harvard.edu/abs/2023Natur.622..707A}
}

@ARTICLE{Goulding2023ApJ...955L..24G,
       author = {{Goulding}, Andy D. and {Greene}, Jenny E. and {Setton}, David J. and {Labbe}, Ivo and {Bezanson}, Rachel and {Miller}, Tim B. and {Atek}, Hakim and {Bogd{\'a}n}, {\'A}kos and {Brammer}, Gabriel and {Chemerynska}, Iryna and {Cutler}, Sam E. and {Dayal}, Pratika and {Fudamoto}, Yoshinobu and {Fujimoto}, Seiji and {Furtak}, Lukas J. and {Kokorev}, Vasily and {Khullar}, Gourav and {Leja}, Joel and {Marchesini}, Danilo and {Natarajan}, Priyamvada and {Nelson}, Erica and {Oesch}, Pascal A. and {Pan}, Richard and {Papovich}, Casey and {Price}, Sedona H. and {van Dokkum}, Pieter and {Wang}, Bingjie and {Weaver}, John R. and {Whitaker}, Katherine E. and {Zitrin}, Adi},
        title = "{UNCOVER: The Growth of the First Massive Black Holes from JWST/NIRSpec-Spectroscopic Redshift Confirmation of an X-Ray Luminous AGN at z = 10.1}",
      journal = {\apjl},
         year = 2023,
        month = sep,
       volume = {955},
       number = {1},
          eid = {L24},
        pages = {L24},
          doi = {10.3847/2041-8213/acf7c5},
archivePrefix = {arXiv},
       eprint = {2308.02750},
 primaryClass = {astro-ph.GA},
       adsurl = {https://ui.adsabs.harvard.edu/abs/2023ApJ...955L..24G}
}

@ARTICLE{Harikane2026arXiv260121833H,
       author = {{Harikane}, Yuichi and {Perez-Gonzalez}, Pablo G. and {Alvarez-Marquez}, Javier and {Ouchi}, Masami and {Nakazato}, Yurina and {Ono}, Yoshiaki and {Nakajima}, Kimihiko and {Umeda}, Hiroya and {Isobe}, Yuki and {Xu}, Yi and {Zhang}, Yechi},
        title = "{A UV-Luminous Galaxy at z=11 with Surprisingly Weak Star Formation Activity}",
      journal = {arXiv e-prints},
         year = 2026,
        month = jan,
          eid = {arXiv:2601.21833},
        pages = {arXiv:2601.21833},
archivePrefix = {arXiv},
       eprint = {2601.21833},
 primaryClass = {astro-ph.GA},
       adsurl = {https://ui.adsabs.harvard.edu/abs/2026arXiv260121833H}
}

@ARTICLE{Korber2026arXiv260119989K,
       author = {{Korber}, Damien and {Schaerer}, Daniel and {Marques-Chaves}, Rui and {Adamo}, Angela and {Basu}, Arghyadeep and {Chisholm}, John and {Dessauges-Zavadsky}, Miroslava and {Kristen. McQuinn}, B.~W. and {Saldana-Lopez}, Alberto and {Atek}, Hakim and {Endsley}, Ryan and {Fujimoto}, Seiji and {Lukas Furtak}, J. and {Kokorev}, Vasily and {Rohan Naidu}, P. and {Pan}, Richard},
        title = "{GLIMPSE-D: Metallicity Decline in Faint Galaxies: Implications for [O III]+Hb Luminosity Function and Reionisation Budget}",
      journal = {arXiv e-prints},
         year = 2026,
        month = jan,
          eid = {arXiv:2601.19989},
        pages = {arXiv:2601.19989},
archivePrefix = {arXiv},
       eprint = {2601.19989},
 primaryClass = {astro-ph.GA},
       adsurl = {https://ui.adsabs.harvard.edu/abs/2026arXiv260119989K}
}

@ARTICLE{Jecmen2026arXiv260119995J,
       author = {{Jecmen}, Michelle C. and {Chisholm}, John and {Atek}, Hakim and {Kokorev}, Vasily and {Endsley}, Ryan and {Chemerynska}, Iryna and {Furtak}, Lukas J. and {Pan}, Richard and {Fujimoto}, Seiji and {Naidu}, Rohan P. and {Mu{\~n}oz}, Julian B. and {Adamo}, Angela and {Asada}, Yoshihisa and {Basu}, Arghyadeep and {Berg}, Danielle A. and {Blaizot}, Jeremy and {Dessauges-Zavadsky}, Miroslava and {Giovinazzo}, Emma and {Hsiao}, Tiger Yu-Yang and {Katz}, Harley and {Korber}, Damien and {McKinney}, Jed and {McQuinn}, Kristen. B.~W. and {Oesch}, Pascal A. and {Schaerer}, Daniel},
        title = "{A GLIMPSE into the UV Continuum Slopes of the Faintest Galaxies in the Epoch of Reionization}",
      journal = {arXiv e-prints},
         year = 2026,
        month = jan,
          eid = {arXiv:2601.19995},
        pages = {arXiv:2601.19995},
archivePrefix = {arXiv},
       eprint = {2601.19995},
 primaryClass = {astro-ph.GA},
       adsurl = {https://ui.adsabs.harvard.edu/abs/2026arXiv260119995J}
}

@ARTICLE{Kerutt2024A&A...684A..42K,
       author = {{Kerutt}, J. and {Oesch}, P.~A. and {Wisotzki}, L. and {Verhamme}, A. and {Atek}, H. and {Herenz}, E.~C. and {Illingworth}, G.~D. and {Kusakabe}, H. and {Matthee}, J. and {Mauerhofer}, V. and {Montes}, M. and {Naidu}, R.~P. and {Nelson}, E. and {Reddy}, N. and {Schaye}, J. and {Simmonds}, C. and {Urrutia}, T. and {Vitte}, E.},
        title = "{Lyman continuum leaker candidates at z {\ensuremath{\sim}} 3-4 in the HDUV based on a spectroscopic sample of MUSE LAEs}",
      journal = {\aap},
         year = 2024,
        month = apr,
       volume = {684},
          eid = {A42},
        pages = {A42},
          doi = {10.1051/0004-6361/202346656},
archivePrefix = {arXiv},
       eprint = {2312.08791},
 primaryClass = {astro-ph.GA},
       adsurl = {https://ui.adsabs.harvard.edu/abs/2024A&A...684A..42K}
}

@ARTICLE{Steidel2018ApJ...869..123S,
       author = {{Steidel}, Charles C. and {Bogosavljevi{\'c}}, Milan and {Shapley}, Alice E. and {Reddy}, Naveen A. and {Rudie}, Gwen C. and {Pettini}, Max and {Trainor}, Ryan F. and {Strom}, Allison L.},
        title = "{The Keck Lyman Continuum Spectroscopic Survey (KLCS): The Emergent Ionizing Spectrum of Galaxies at z {\ensuremath{\sim}} 3}",
      journal = {\apj},
         year = 2018,
        month = dec,
       volume = {869},
       number = {2},
          eid = {123},
        pages = {123},
          doi = {10.3847/1538-4357/aaed28},
archivePrefix = {arXiv},
       eprint = {1805.06071},
 primaryClass = {astro-ph.GA},
       adsurl = {https://ui.adsabs.harvard.edu/abs/2018ApJ...869..123S}
}

@ARTICLE{Marques-Chaves2021MNRAS.507..524M,
       author = {{Marques-Chaves}, R. and {Schaerer}, D. and {{\'A}lvarez-M{\'a}rquez}, J. and {Colina}, L. and {Dessauges-Zavadsky}, M. and {P{\'e}rez-Fournon}, I. and {Saldana-Lopez}, A. and {Verhamme}, A.},
        title = "{The UV-brightest Lyman continuum emitting star-forming galaxy}",
      journal = {\mnras},
         year = 2021,
        month = oct,
       volume = {507},
       number = {1},
        pages = {524-538},
          doi = {10.1093/mnras/stab2187},
archivePrefix = {arXiv},
       eprint = {2107.12313},
 primaryClass = {astro-ph.GA},
       adsurl = {https://ui.adsabs.harvard.edu/abs/2021MNRAS.507..524M}
}

@software{Newville2025zndo..15014437N,
       author = {{Newville}, Matthew and {Otten}, Renee and {Nelson}, Andrew and {Stensitzki}, Till and {Ingargiola}, Antonino and {Allan}, Daniel and {Fox}, Austin and {Carter}, Faustin and {Rawlik}, Michal},
        title = "{LMFIT: Non-Linear Least-Squares Minimization and Curve-Fitting for Python}",
         year = 2025,
        month = mar,
          eid = {10.5281/zenodo.15014437},
          doi = {10.5281/zenodo.15014437},
      version = {1.3.3},
    publisher = {Zenodo},
       adsurl = {https://ui.adsabs.harvard.edu/abs/2025zndo..15014437N}
}

@ARTICLE{Chemerynska2026MNRAS.tmp...12C,
       author = {{Chemerynska}, Iryna and {Atek}, Hakim and {Furtak}, Lukas J. and {Chisholm}, John and {Endsley}, Ryan and {Kokorev}, Vasily and {Rosdahl}, Joki and {Blaizot}, Jeremy and {Adamo}, Angela and {Bouwens}, Rychard and {Fujimoto}, Seiji and {Korber}, Damien and {Mason}, Charlotte and {McQuinn}, Kristen B.~W. and {Mu{\~n}oz}, Julian B. and {Natarajan}, Priyamvada and {Nelson}, Erica and {Oesch}, Pascal A. and {Pan}, Richard and {Richard}, Johan and {Saldana-Lopez}, Alberto and {Schaerer}, Daniel and {Volonteri}, Marta and {Zitrin}, Adi and {Berg}, Danielle A. and {Claeyssens}, Ad{\'e}la{\"\i}de and {Dessauges-Zavadsky}, Miroslava and {Jecmen}, Michelle and {Labb{\'e}}, Ivo and {Naidu}, Rohan and {Trebitsch}, Maxime},
        title = "{The first GLIMPSE of the faint galaxy population at Cosmic Dawn with JWST: The evolution of the ultraviolet luminosity function across z \raisebox{-0.5ex}\textasciitilde 9 - 15}",
      journal = {\mnras},
         year = 2026,
        month = jan,
          doi = {10.1093/mnras/staf2267},
archivePrefix = {arXiv},
       eprint = {2509.24881},
 primaryClass = {astro-ph.GA},
       adsurl = {https://ui.adsabs.harvard.edu/abs/2026MNRAS.tmp...12C}
}

@ARTICLE{Heintz2024Sci...384..890H,
       author = {{Heintz}, Kasper E. and {Watson}, Darach and {Brammer}, Gabriel and {Vejlgaard}, Simone and {Hutter}, Anne and {Strait}, Victoria B. and {Matthee}, Jorryt and {Oesch}, Pascal A. and {Jakobsson}, P{\'a}ll and {Tanvir}, Nial R. and {Laursen}, Peter and {Naidu}, Rohan P. and {Mason}, Charlotte A. and {Killi}, Meghana and {Jung}, Intae and {Hsiao}, Tiger Yu-Yang and {Abdurro'uf} and {Coe}, Dan and {Arrabal Haro}, Pablo and {Finkelstein}, Steven L. and {Toft}, Sune},
        title = "{Strong damped Lyman-{\ensuremath{\alpha}} absorption in young star-forming galaxies at redshifts 9 to 11}",
      journal = {Science},
         year = 2024,
        month = may,
       volume = {384},
       number = {6698},
        pages = {890-894},
          doi = {10.1126/science.adj0343},
archivePrefix = {arXiv},
       eprint = {2306.00647},
 primaryClass = {astro-ph.GA},
       adsurl = {https://ui.adsabs.harvard.edu/abs/2024Sci...384..890H}
}

@ARTICLE{Burgarella2025A&A...699A.336B,
       author = {{Burgarella}, Denis and {Buat}, V{\'e}ronique and {Theul{\'e}}, Patrice and {Zavala}, Jorge and {Dickinson}, Mark and {Arrabal Haro}, Pablo and {Bagley}, Micaela B. and {Boquien}, M{\'e}d{\'e}ric and {Cleri}, Nikko and {Dewachter}, Tim and {Ferguson}, Henry C. and {Fern{\`a}ndez}, Vital and {Finkelstein}, Steven L. and {Gawiser}, Eric and {Grazian}, Andrea and {Grogin}, Norman and {Holwerda}, Benne W. and {Kartaltepe}, Jeyhan S. and {Kewley}, Lisa and {Kirkpatrick}, Allison and {Kocevski}, Dale and {Koekemoer}, Anton M. and {Long}, Arianna and {Lotz}, Jennifer and {Lucas}, Ray A. and {Mobasher}, Bahram and {Papovich}, Casey and {P{\'e}rez-Gonz{\`a}lez}, Pablo G. and {Pirzkal}, Nor and {Ravindranath}, Swara and {Rodighiero}, Giulia and {Roehlly}, Yannick and {Rose}, Caitlin and {Seill{\'e}}, Lise-Marie and {Somerville}, Rachel and {Wilkins}, Steve and {Yang}, Guang and {Yung}, L.~Y. Aaron},
        title = "{CEERS: Possibly forging the first dust grains in the universe: A population of galaxies with spectroscopically derived extremely low dust attenuation (GELDA) at 4.0 < z {\ensuremath{\lesssim}} 11.4}",
      journal = {\aap},
         year = 2025,
        month = jul,
       volume = {699},
          eid = {A336},
        pages = {A336},
          doi = {10.1051/0004-6361/202554231},
archivePrefix = {arXiv},
       eprint = {2504.13118},
 primaryClass = {astro-ph.GA},
       adsurl = {https://ui.adsabs.harvard.edu/abs/2025A&A...699A.336B}
}

@ARTICLE{Donnan2026arXiv260111515D,
       author = {{Donnan}, Callum T. and {McLeod}, Derek J. and {McLure}, Ross J. and {Dunlop}, James S. and {Cullen}, Fergus and {Dickinson}, Mark and {Arrabal Haro}, Pablo and {Taylor}, Anthony J. and {Bondestam}, Cecilia and {Liu}, Feng-Yuan and {Arellano-C{\'o}rdova}, Karla Z. and {Barrufet}, Laia and {Begley}, Ryan and {Carnall}, Adam C. and {Golawska}, Hanna and {Leung}, Ho-Hin and {Scholte}, Dirk and {Stanton}, Thomas M.},
        title = "{Spectroscopic confirmation of a large and luminous galaxy with weak emission lines at $\mathbf{z = 13.53}$}",
      journal = {arXiv e-prints},
         year = 2026,
        month = jan,
          eid = {arXiv:2601.11515},
        pages = {arXiv:2601.11515},
          doi = {10.48550/arXiv.2601.11515},
archivePrefix = {arXiv},
       eprint = {2601.11515},
 primaryClass = {astro-ph.GA},
       adsurl = {https://ui.adsabs.harvard.edu/abs/2026arXiv260111515D}
}

@ARTICLE{Reddy2023ApJ...951...56R,
       author = {{Reddy}, Naveen A. and {Sanders}, Ryan L. and {Shapley}, Alice E. and {Topping}, Michael W. and {Kriek}, Mariska and {Coil}, Alison L. and {Mobasher}, Bahram and {Siana}, Brian and {Rezaee}, Saeed},
        title = "{The Impact of Star-formation-rate Surface Density on the Electron Density and Ionization Parameter of High-redshift Galaxies}",
      journal = {\apj},
         year = 2023,
        month = jul,
       volume = {951},
       number = {1},
          eid = {56},
        pages = {56},
          doi = {10.3847/1538-4357/acd0b1},
archivePrefix = {arXiv},
       eprint = {2302.10213},
 primaryClass = {astro-ph.GA},
       adsurl = {https://ui.adsabs.harvard.edu/abs/2023ApJ...951...56R}
}

@ARTICLE{Reddy2023ApJ...952..167R,
       author = {{Reddy}, Naveen A. and {Topping}, Michael W. and {Sanders}, Ryan L. and {Shapley}, Alice E. and {Brammer}, Gabriel},
        title = "{A JWST/NIRSpec Exploration of the Connection between Ionization Parameter, Electron Density, and Star-formation-rate Surface Density in z = 2.7-6.3 Galaxies}",
      journal = {\apj},
         year = 2023,
        month = aug,
       volume = {952},
       number = {2},
          eid = {167},
        pages = {167},
          doi = {10.3847/1538-4357/acd754},
archivePrefix = {arXiv},
       eprint = {2303.11397},
 primaryClass = {astro-ph.GA},
       adsurl = {https://ui.adsabs.harvard.edu/abs/2023ApJ...952..167R}
}

@ARTICLE{Messa2025A&A...694A..59M,
       author = {{Messa}, M. and {Vanzella}, E. and {Loiacono}, F. and {Bergamini}, P. and {Castellano}, M. and {Sun}, B. and {Willott}, C. and {Windhorst}, R.~A. and {Yan}, H. and {Angora}, G. and {Rosati}, P. and {Adamo}, A. and {Annibali}, F. and {Bolamperti}, A. and {Brada{\v{c}}}, M. and {Bradley}, L.~D. and {Calura}, F. and {Claeyssens}, A. and {Comastri}, A. and {Conselice}, C.~J. and {D'Silva}, J.~C.~J. and {Dickinson}, M. and {Frye}, B.~L. and {Grillo}, C. and {Grogin}, N.~A. and {Gruppioni}, C. and {Koekemoer}, A.~M. and {Meneghetti}, M. and {Me{\v{s}}tri{\'c}}, U. and {Pascale}, R. and {Ravindranath}, S. and {Ricotti}, M. and {Summers}, J. and {Zanella}, A.},
        title = "{Anatomy of a z = 6 Lyman-{\ensuremath{\alpha}} emitter down to parsec scales: Extreme UV slopes, metal-poor regions, and possibly leaking star clusters}",
      journal = {\aap},
         year = 2025,
        month = feb,
       volume = {694},
          eid = {A59},
        pages = {A59},
          doi = {10.1051/0004-6361/202451695},
archivePrefix = {arXiv},
       eprint = {2407.20331},
 primaryClass = {astro-ph.GA},
       adsurl = {https://ui.adsabs.harvard.edu/abs/2025A&A...694A..59M}
}

@ARTICLE{Algera2025arXiv251214486A,
       author = {{Algera}, Hiddo S.~B. and {Weaver}, John R. and {Bakx}, Tom J.~L.~C. and {Aravena}, Manuel and {Bouwens}, Rychard J. and {Cescon}, Karin and {Chen}, Chian-Chou and {da Cunha}, Elisabete and {Dayal}, Pratika and {Faisst}, Andreas and {Ferrara}, Andrea and {Fujimoto}, Seiji and {Hashimoto}, Takuya and {Heintz}, Kasper and {Herrera-Camus}, Rodrigo and {Hodge}, Jacqueline and {Inami}, Hanae and {Inoue}, Akio K. and {Matthee}, Jorryt and {Meyer}, Romain and {Mizukoshi}, Shoichiro and {Mondal}, Chayan and {Nanayakkara}, Themiya and {Oesch}, Pascal A. and {Pallottini}, Andrea and {R{\"o}ttgering}, Huub and {Rowland}, Lucie E. and {Schouws}, Sander and {Smit}, Renske and {Sommovigo}, Laura and {Stark}, Daniel P. and {Sugahara}, Yuma and {Vallini}, Livia and {Vijarnwannaluk}, Bovornpratch and {van der Werf}, Paul and {Werner}, Norbert and {Witstok}, Joris and {Xiao}, Mengyuan},
        title = "{A first systematic study of [OIII] 88$μ$m at $z>8$: two luminous oxygen lines and a powerful ionized outflow in the first 600 million years}",
      journal = {arXiv e-prints},
         year = 2025,
        month = dec,
          eid = {arXiv:2512.14486},
        pages = {arXiv:2512.14486},
          doi = {10.48550/arXiv.2512.14486},
archivePrefix = {arXiv},
       eprint = {2512.14486},
 primaryClass = {astro-ph.GA},
       adsurl = {https://ui.adsabs.harvard.edu/abs/2025arXiv251214486A}
}

@ARTICLE{Napolitano2025A&A...693A..50N,
       author = {{Napolitano}, L. and {Castellano}, M. and {Pentericci}, L. and {Arrabal Haro}, P. and {Fontana}, A. and {Treu}, T. and {Bergamini}, P. and {Calabr{\`o}}, A. and {Mascia}, S. and {Morishita}, T. and {Roberts-Borsani}, G. and {Santini}, P. and {Vanzella}, E. and {Vulcani}, B. and {Zakharova}, D. and {Bakx}, T. and {Dickinson}, M. and {Grillo}, C. and {Leethochawalit}, N. and {Llerena}, M. and {Merlin}, E. and {Paris}, D. and {Rojas-Ruiz}, S. and {Rosati}, P. and {Wang}, X. and {Yoon}, I. and {Zavala}, J.},
        title = "{Seven wonders of Cosmic Dawn: JWST confirms a high abundance of galaxies and AGN at z ≃ 9{\textendash}11 in the GLASS field}",
      journal = {\aap},
         year = 2025,
        month = jan,
       volume = {693},
          eid = {A50},
        pages = {A50},
          doi = {10.1051/0004-6361/202452090},
archivePrefix = {arXiv},
       eprint = {2410.10967},
 primaryClass = {astro-ph.GA},
       adsurl = {https://ui.adsabs.harvard.edu/abs/2025A&A...693A..50N}
}

@ARTICLE{Harikane2025ApJ...980..138H,
       author = {{Harikane}, Yuichi and {Inoue}, Akio K. and {Ellis}, Richard S. and {Ouchi}, Masami and {Nakazato}, Yurina and {Yoshida}, Naoki and {Ono}, Yoshiaki and {Sun}, Fengwu and {Sato}, Riku A. and {Ferrami}, Giovanni and {Fujimoto}, Seiji and {Kashikawa}, Nobunari and {McLeod}, Derek J. and {P{\'e}rez-Gonz{\'a}lez}, Pablo G. and {Sawicki}, Marcin and {Sugahara}, Yuma and {Xu}, Yi and {Yamanaka}, Satoshi and {Carnall}, Adam C. and {Cullen}, Fergus and {Dunlop}, James S. and {Egami}, Eiichi and {Grogin}, Norman and {Isobe}, Yuki and {Koekemoer}, Anton M. and {Laporte}, Nicolas and {Lee}, Chien-Hsiu and {Magee}, Dan and {Matsuo}, Hiroshi and {Matsuoka}, Yoshiki and {Mawatari}, Ken and {Nakajima}, Kimihiko and {Nakane}, Minami and {Tamura}, Yoichi and {Umeda}, Hiroya and {Yanagisawa}, Hiroto},
        title = "{JWST, ALMA, and Keck Spectroscopic Constraints on the UV Luminosity Functions at z {\ensuremath{\sim}} 7{\textendash}14: Clumpiness and Compactness of the Brightest Galaxies in the Early Universe}",
      journal = {\apj},
         year = 2025,
        month = feb,
       volume = {980},
       number = {1},
          eid = {138},
        pages = {138},
          doi = {10.3847/1538-4357/ad9b2c},
archivePrefix = {arXiv},
       eprint = {2406.18352},
 primaryClass = {astro-ph.GA},
       adsurl = {https://ui.adsabs.harvard.edu/abs/2025ApJ...980..138H}
}

@ARTICLE{Harikane2024ApJ...960...56H,
       author = {{Harikane}, Yuichi and {Nakajima}, Kimihiko and {Ouchi}, Masami and {Umeda}, Hiroya and {Isobe}, Yuki and {Ono}, Yoshiaki and {Xu}, Yi and {Zhang}, Yechi},
        title = "{Pure Spectroscopic Constraints on UV Luminosity Functions and Cosmic Star Formation History from 25 Galaxies at z $_{spec}$ = 8.61-13.20 Confirmed with JWST/NIRSpec}",
      journal = {\apj},
         year = 2024,
        month = jan,
       volume = {960},
       number = {1},
          eid = {56},
        pages = {56},
          doi = {10.3847/1538-4357/ad0b7e},
archivePrefix = {arXiv},
       eprint = {2304.06658},
 primaryClass = {astro-ph.GA},
       adsurl = {https://ui.adsabs.harvard.edu/abs/2024ApJ...960...56H}
}

@ARTICLE{Castellano2022ApJ...938L..15C,
       author = {{Castellano}, Marco and {Fontana}, Adriano and {Treu}, Tommaso and {Santini}, Paola and {Merlin}, Emiliano and {Leethochawalit}, Nicha and {Trenti}, Michele and {Vanzella}, Eros and {Mestric}, Uros and {Bonchi}, Andrea and {Belfiori}, Davide and {Nonino}, Mario and {Paris}, Diego and {Polenta}, Gianluca and {Roberts-Borsani}, Guido and {Boyett}, Kristan and {Brada{\v{c}}}, Maru{\v{s}}a and {Calabr{\`o}}, Antonello and {Glazebrook}, Karl and {Grillo}, Claudio and {Mascia}, Sara and {Mason}, Charlotte and {Mercurio}, Amata and {Morishita}, Takahiro and {Nanayakkara}, Themiya and {Pentericci}, Laura and {Rosati}, Piero and {Vulcani}, Benedetta and {Wang}, Xin and {Yang}, Lilan},
        title = "{Early Results from GLASS-JWST. III. Galaxy Candidates at z  9-15}",
      journal = {\apjl},
         year = 2022,
        month = oct,
       volume = {938},
       number = {2},
          eid = {L15},
        pages = {L15},
          doi = {10.3847/2041-8213/ac94d0},
archivePrefix = {arXiv},
       eprint = {2207.09436},
 primaryClass = {astro-ph.GA},
       adsurl = {https://ui.adsabs.harvard.edu/abs/2022ApJ...938L..15C}
}

@ARTICLE{Bergvall2013A&A...554A..38B,
       author = {{Bergvall}, N. and {Leitet}, E. and {Zackrisson}, E. and {Marquart}, T.},
        title = "{Lyman continuum leaking galaxies. Search strategies and local candidates}",
      journal = {\aap},
         year = 2013,
        month = jun,
       volume = {554},
          eid = {A38},
        pages = {A38},
          doi = {10.1051/0004-6361/201118433},
archivePrefix = {arXiv},
       eprint = {1303.1576},
 primaryClass = {astro-ph.CO},
       adsurl = {https://ui.adsabs.harvard.edu/abs/2013A&A...554A..38B}
}

@ARTICLE{Ferrara2025OJAp....8E.125F,
       author = {{Ferrara}, Andrea and {Giavalisco}, M. and {Pentericci}, L. and {Vanzella}, E. and {Calabr{\`o}}, A. and {Llerena}, M.},
        title = "{Redshift evolution of Lyman continuum escape fraction after JWST}",
      journal = {The Open Journal of Astrophysics},
         year = 2025,
        month = aug,
       volume = {8},
          eid = {125},
        pages = {125},
          doi = {10.33232/001c.143600},
archivePrefix = {arXiv},
       eprint = {2505.10619},
 primaryClass = {astro-ph.GA},
       adsurl = {https://ui.adsabs.harvard.edu/abs/2025OJAp....8E.125F}
}

@ARTICLE{Saxena2024arXiv241114532S,
       author = {{Saxena}, Aayush and {Cameron}, Alex J. and {Katz}, Harley and {Bunker}, Andrew J. and {Chevallard}, Jacopo and {D'Eugenio}, Francesco and {Arribas}, Santiago and {Bhatawdekar}, Rachana and {Boyett}, Kristan and {Cargile}, Phillip A. and {Carniani}, Stefano and {Charlot}, Stephane and {Curti}, Mirko and {Curtis-Lake}, Emma and {Hainline}, Kevin and {Ji}, Zhiyuan and {Johnson}, Benjamin D. and {Jones}, Gareth C. and {Kumari}, Nimisha and {Laseter}, Isaac and {Maseda}, Michael V. and {Robertson}, Brant and {Simmonds}, Charlotte and {Tacchella}, Sandro and {Ubler}, Hannah and {Williams}, Christina C. and {Willott}, Chris and {Witstok}, Joris and {Zhu}, Yongda},
        title = "{Hitting the slopes: A spectroscopic view of UV continuum slopes of galaxies reveals a reddening at z > 9.5}",
      journal = {arXiv e-prints},
         year = 2024,
        month = nov,
          eid = {arXiv:2411.14532},
        pages = {arXiv:2411.14532},
          doi = {10.48550/arXiv.2411.14532},
archivePrefix = {arXiv},
       eprint = {2411.14532},
 primaryClass = {astro-ph.GA},
       adsurl = {https://ui.adsabs.harvard.edu/abs/2024arXiv241114532S}
}

@ARTICLE{Boyett2024NatAs...8..657B,
       author = {{Boyett}, Kristan and {Trenti}, Michele and {Leethochawalit}, Nicha and {Calabr{\'o}}, Antonello and {Metha}, Benjamin and {Roberts-Borsani}, Guido and {Dalmasso}, Nicol{\'o} and {Yang}, Lilan and {Santini}, Paola and {Treu}, Tommaso and {Jones}, Tucker and {Henry}, Alaina and {Mason}, Charlotte A. and {Morishita}, Takahiro and {Nanayakkara}, Themiya and {Roy}, Namrata and {Wang}, Xin and {Fontana}, Adriano and {Merlin}, Emiliano and {Castellano}, Marco and {Paris}, Diego and {Brada{\v{c}}}, Maru{\v{s}}a and {Malkan}, Matt and {Marchesini}, Danilo and {Mascia}, Sara and {Glazebrook}, Karl and {Pentericci}, Laura and {Vanzella}, Eros and {Vulcani}, Benedetta},
        title = "{A massive interacting galaxy 510 million years after the Big Bang}",
      journal = {Nature Astronomy},
         year = 2024,
        month = may,
       volume = {8},
        pages = {657-672},
          doi = {10.1038/s41550-024-02218-7},
archivePrefix = {arXiv},
       eprint = {2303.00306},
 primaryClass = {astro-ph.GA},
       adsurl = {https://ui.adsabs.harvard.edu/abs/2024NatAs...8..657B}
}

@ARTICLE{Helton2025arXiv251219695H,
       author = {{Helton}, Jakob M. and {Morrison}, Jane E. and {Hainline}, Kevin N. and {D'Eugenio}, Francesco and {Rieke}, George H. and {Alberts}, Stacey and {Carniani}, Stefano and {Leja}, Joel and {Li}, Yijia and {Rinaldi}, Pierluigi and {Scholtz}, Jan and {Stone}, Meredith and {Willmer}, Christopher N.~A. and {Wu}, Zihao and {Baker}, William M. and {Bunker}, Andrew J. and {Charlot}, Stephane and {Chevallard}, Jacopo and {Cleri}, Nikko J. and {Curti}, Mirko and {Curtis-Lake}, Emma and {Egami}, Eiichi and {Eisenstein}, Daniel J. and {Jakobsen}, Peter and {Ji}, Zhiyuan and {Johnson}, Benjamin D. and {Kumari}, Nimisha and {Lin}, Xiaojing and {Lyu}, Jianwei and {Maiolino}, Roberto and {Maseda}, Michael and {P{\'e}rez-Gonz{\'a}lez}, Pablo G. and {Rieke}, Marcia J. and {Robertson}, Brant and {Saxena}, Aayush and {Sun}, Fengwu and {Tacchella}, Sandro and {{\"U}bler}, Hannah and {Venturi}, Giacomo and {Williams}, Christina C. and {Willott}, Chris and {Witstok}, Joris and {Zhu}, Yongda},
        title = "{Ionizing Photon Production Efficiencies and Chemical Abundances at Cosmic Dawn Revealed by Ultra-Deep Rest-Frame Optical Spectroscopy of JADES-GS-z14-0}",
      journal = {arXiv e-prints},
         year = 2025,
        month = dec,
          eid = {arXiv:2512.19695},
        pages = {arXiv:2512.19695},
          doi = {10.48550/arXiv.2512.19695},
archivePrefix = {arXiv},
       eprint = {2512.19695},
 primaryClass = {astro-ph.GA},
       adsurl = {https://ui.adsabs.harvard.edu/abs/2025arXiv251219695H}
}

@ARTICLE{Donnan2025ApJ...993..224D,
       author = {{Donnan}, Callum T. and {Dickinson}, Mark and {Taylor}, Anthony J. and {Arrabal Haro}, Pablo and {Finkelstein}, Steven L. and {Stanton}, Thomas M. and {Jung}, Intae and {Papovich}, Casey and {Akins}, Hollis B. and {Koekemoer}, Anton M. and {McLeod}, Derek J. and {Napolitano}, Lorenzo and {Amor{\'\i}n}, Ricardo O. and {Begley}, Ryan and {Burgarella}, Denis and {Carnall}, Adam C. and {Casey}, Caitlin M. and {Calabr{\`o}}, Antonello and {Cullen}, Fergus and {Dunlop}, James S. and {Ellis}, Richard S. and {Fern{\'a}ndez}, Vital and {Giavalisco}, Mauro and {Hirschmann}, Michaela and {Hu}, Weida and {Illingworth}, Garth and {Kartaltepe}, Jeyhan S. and {Kocevski}, Dale D. and {Kokorev}, Vasily and {Leung}, Ho-Hin and {Lucas}, Ray A. and {Morales}, Alexa M. and {McLure}, Ross and {Pentericci}, Laura and {P{\'e}rez-Gonz{\'a}lez}, Pablo G. and {Somerville}, Rachel S. and {Stevenson}, Struan and {Trump}, Jonathan R. and {Yung}, L.~Y. Aaron and {Zavala}, Jorge A.},
        title = "{Very Bright, Very Blue, and Very Red: JWST CAPERS Analysis of Highly Luminous Galaxies with Extreme Ultraviolet Slopes at z = 10}",
      journal = {\apj},
         year = 2025,
        month = nov,
       volume = {993},
       number = {2},
          eid = {224},
        pages = {224},
          doi = {10.3847/1538-4357/ae0a1f},
archivePrefix = {arXiv},
       eprint = {2507.10518},
 primaryClass = {astro-ph.GA},
       adsurl = {https://ui.adsabs.harvard.edu/abs/2025ApJ...993..224D}
}

@ARTICLE{Atek2023MNRAS.524.5486A,
       author = {{Atek}, Hakim and {Chemerynska}, Iryna and {Wang}, Bingjie and {Furtak}, Lukas J. and {Weibel}, Andrea and {Oesch}, Pascal and {Weaver}, John R. and {Labb{\'e}}, Ivo and {Bezanson}, Rachel and {van Dokkum}, Pieter and {Zitrin}, Adi and {Dayal}, Pratika and {Williams}, Christina C. and {Nannayakkara}, Themiya and {Price}, Sedona H. and {Brammer}, Gabriel and {Goulding}, Andy D. and {Leja}, Joel and {Marchesini}, Danilo and {Nelson}, Erica J. and {Pan}, Richard and {Whitaker}, Katherine E.},
        title = "{JWST UNCOVER: discovery of z > 9 galaxy candidates behind the lensing cluster Abell 2744}",
      journal = {\mnras},
         year = 2023,
        month = oct,
       volume = {524},
       number = {4},
        pages = {5486-5496},
          doi = {10.1093/mnras/stad1998},
archivePrefix = {arXiv},
       eprint = {2305.01793},
 primaryClass = {astro-ph.GA},
       adsurl = {https://ui.adsabs.harvard.edu/abs/2023MNRAS.524.5486A}
}

@ARTICLE{Bouwens2010ApJ...708L..69B,
       author = {{Bouwens}, R.~J. and {Illingworth}, G.~D. and {Oesch}, P.~A. and {Trenti}, M. and {Stiavelli}, M. and {Carollo}, C.~M. and {Franx}, M. and {van Dokkum}, P.~G. and {Labb{\'e}}, I. and {Magee}, D.},
        title = "{Very Blue UV-Continuum Slope {\ensuremath{\beta}} of Low Luminosity z \raisebox{-0.5ex}\textasciitilde 7 Galaxies from WFC3/IR: Evidence for Extremely Low Metallicities?}",
      journal = {\apjl},
         year = 2010,
        month = jan,
       volume = {708},
       number = {2},
        pages = {L69-L73},
          doi = {10.1088/2041-8205/708/2/L69},
archivePrefix = {arXiv},
       eprint = {0910.0001},
 primaryClass = {astro-ph.CO},
       adsurl = {https://ui.adsabs.harvard.edu/abs/2010ApJ...708L..69B}
}

@ARTICLE{Morales2024ApJ...964L..24M,
       author = {{Morales}, Alexa M. and {Finkelstein}, Steven L. and {Leung}, Gene C.~K. and {Bagley}, Micaela B. and {Cleri}, Nikko J. and {Dave}, Romeel and {Dickinson}, Mark and {Ferguson}, Henry C. and {Hathi}, Nimish P. and {Jones}, Ewan and {Koekemoer}, Anton M. and {Papovich}, Casey and {P{\'e}rez-Gonz{\'a}lez}, Pablo G. and {Pirzkal}, Nor and {Smith}, Britton and {Wilkins}, Stephen M. and {Yung}, L.~Y. Aaron},
        title = "{Rest-frame UV Colors for Faint Galaxies at z {\ensuremath{\sim}} 9{\textendash}16 with the JWST NGDEEP Survey}",
      journal = {\apjl},
         year = 2024,
        month = apr,
       volume = {964},
       number = {2},
          eid = {L24},
        pages = {L24},
          doi = {10.3847/2041-8213/ad2de4},
archivePrefix = {arXiv},
       eprint = {2311.04294},
 primaryClass = {astro-ph.GA},
       adsurl = {https://ui.adsabs.harvard.edu/abs/2024ApJ...964L..24M}
}

@ARTICLE{Topping2024MNRAS.529.4087T,
       author = {{Topping}, Michael W. and {Stark}, Daniel P. and {Endsley}, Ryan and {Whitler}, Lily and {Hainline}, Kevin and {Johnson}, Benjamin D. and {Robertson}, Brant and {Tacchella}, Sandro and {Chen}, Zuyi and {Alberts}, Stacey and {Baker}, William M. and {Bunker}, Andrew J. and {Carniani}, Stefano and {Charlot}, Stephane and {Chevallard}, Jacopo and {Curtis-Lake}, Emma and {DeCoursey}, Christa and {Egami}, Eiichi and {Eisenstein}, Daniel J. and {Ji}, Zhiyuan and {Maiolino}, Roberto and {Williams}, Christina C. and {Willmer}, Christopher N.~A. and {Willott}, Chris and {Witstok}, Joris},
        title = "{The UV continuum slopes of early star-forming galaxies in JADES}",
      journal = {\mnras},
         year = 2024,
        month = apr,
       volume = {529},
       number = {4},
        pages = {4087-4103},
          doi = {10.1093/mnras/stae800},
archivePrefix = {arXiv},
       eprint = {2307.08835},
 primaryClass = {astro-ph.GA},
       adsurl = {https://ui.adsabs.harvard.edu/abs/2024MNRAS.529.4087T}
}

@ARTICLE{Bhatawdekar2021ApJ...909..144B,
       author = {{Bhatawdekar}, Rachana and {Conselice}, Christopher J.},
        title = "{UV Spectral Slopes at z = 6-9 in the Hubble Frontier Fields: Lack of Evidence for Unusual or Population III Stellar Populations}",
      journal = {\apj},
         year = 2021,
        month = mar,
       volume = {909},
       number = {2},
          eid = {144},
        pages = {144},
          doi = {10.3847/1538-4357/abdd3f},
archivePrefix = {arXiv},
       eprint = {2006.00013},
 primaryClass = {astro-ph.GA},
       adsurl = {https://ui.adsabs.harvard.edu/abs/2021ApJ...909..144B}
}

@ARTICLE{Bouwens2012ApJ...754...83B,
       author = {{Bouwens}, R.~J. and {Illingworth}, G.~D. and {Oesch}, P.~A. and {Franx}, M. and {Labb{\'e}}, I. and {Trenti}, M. and {van Dokkum}, P. and {Carollo}, C.~M. and {Gonz{\'a}lez}, V. and {Smit}, R. and {Magee}, D.},
        title = "{UV-continuum Slopes at z \raisebox{-0.5ex}\textasciitilde 4-7 from the HUDF09+ERS+CANDELS Observations: Discovery of a Well-defined UV Color-Magnitude Relationship for z >= 4 Star-forming Galaxies}",
      journal = {\apj},
         year = 2012,
        month = aug,
       volume = {754},
       number = {2},
          eid = {83},
        pages = {83},
          doi = {10.1088/0004-637X/754/2/83},
archivePrefix = {arXiv},
       eprint = {1109.0994},
 primaryClass = {astro-ph.CO},
       adsurl = {https://ui.adsabs.harvard.edu/abs/2012ApJ...754...83B}
}

@ARTICLE{Bouwens2014ApJ...793..115B,
       author = {{Bouwens}, R.~J. and {Illingworth}, G.~D. and {Oesch}, P.~A. and {Labb{\'e}}, I. and {van Dokkum}, P.~G. and {Trenti}, M. and {Franx}, M. and {Smit}, R. and {Gonzalez}, V. and {Magee}, D.},
        title = "{UV-continuum Slopes of >4000 z \raisebox{-0.5ex}\textasciitilde 4-8 Galaxies from the HUDF/XDF, HUDF09, ERS, CANDELS-South, and CANDELS-North Fields}",
      journal = {\apj},
         year = 2014,
        month = oct,
       volume = {793},
       number = {2},
          eid = {115},
        pages = {115},
          doi = {10.1088/0004-637X/793/2/115},
archivePrefix = {arXiv},
       eprint = {1306.2950},
 primaryClass = {astro-ph.CO},
       adsurl = {https://ui.adsabs.harvard.edu/abs/2014ApJ...793..115B}
}

@ARTICLE{Finkelstein2012ApJ...756..164F,
       author = {{Finkelstein}, Steven L. and {Papovich}, Casey and {Salmon}, Brett and {Finlator}, Kristian and {Dickinson}, Mark and {Ferguson}, Henry C. and {Giavalisco}, Mauro and {Koekemoer}, Anton M. and {Reddy}, Naveen A. and {Bassett}, Robert and {Conselice}, Christopher J. and {Dunlop}, James S. and {Faber}, S.~M. and {Grogin}, Norman A. and {Hathi}, Nimish P. and {Kocevski}, Dale D. and {Lai}, Kamson and {Lee}, Kyoung-Soo and {McLure}, Ross J. and {Mobasher}, Bahram and {Newman}, Jeffrey A.},
        title = "{Candels: The Evolution of Galaxy Rest-frame Ultraviolet Colors from z = 8 to 4}",
      journal = {\apj},
         year = 2012,
        month = sep,
       volume = {756},
       number = {2},
          eid = {164},
        pages = {164},
          doi = {10.1088/0004-637X/756/2/164},
archivePrefix = {arXiv},
       eprint = {1110.3785},
 primaryClass = {astro-ph.CO},
       adsurl = {https://ui.adsabs.harvard.edu/abs/2012ApJ...756..164F}
}

@ARTICLE{Tang2025arXiv250708245T,
       author = {{Tang}, Mengtao and {Stark}, Daniel P. and {Mason}, Charlotte A. and {Gelli}, Viola and {Chen}, Zuyi and {Topping}, Michael W.},
        title = "{The JWST Spectroscopic Properties of Galaxies at $z=9-14$}",
      journal = {arXiv e-prints},
         year = 2025,
        month = jul,
          eid = {arXiv:2507.08245},
        pages = {arXiv:2507.08245},
          doi = {10.48550/arXiv.2507.08245},
archivePrefix = {arXiv},
       eprint = {2507.08245},
 primaryClass = {astro-ph.GA},
       adsurl = {https://ui.adsabs.harvard.edu/abs/2025arXiv250708245T}
}

@ARTICLE{Roberts-Borsani2025arXiv250821708R,
       author = {{Roberts-Borsani}, Guido and {Oesch}, Pascal and {Ellis}, Richard and {Weibel}, Andrea and {Giovinazzo}, Emma and {Bouwens}, Rychard and {Dayal}, Pratika and {Fontana}, Adriano and {Heintz}, Kasper and {Matthee}, Jorryt and {Meyer}, Romain and {Pentericci}, Laura and {Shapley}, Alice and {Tacchella}, Sandro and {Treu}, Tommaso and {Walter}, Fabian and {Atek}, Hakim and {Bose}, Sownak and {Castellano}, Marco and {Fudamoto}, Yoshinobu and {Morishita}, Takahiro and {Naidu}, Rohan and {Sanders}, Ryan and {van der Wel}, Arjen},
        title = "{JWST Spectroscopic Insights Into the Diversity of Galaxies in the First 500 Myr: Short-Lived Snapshots Along a Common Evolutionary Pathway}",
      journal = {arXiv e-prints},
         year = 2025,
        month = aug,
          eid = {arXiv:2508.21708},
        pages = {arXiv:2508.21708},
          doi = {10.48550/arXiv.2508.21708},
archivePrefix = {arXiv},
       eprint = {2508.21708},
 primaryClass = {astro-ph.GA},
       adsurl = {https://ui.adsabs.harvard.edu/abs/2025arXiv250821708R}
}

@ARTICLE{Naidu2025arXiv250511263N,
       author = {{Naidu}, Rohan P. and {Oesch}, Pascal A. and {Brammer}, Gabriel and {Weibel}, Andrea and {Li}, Yijia and {Matthee}, Jorryt and {Chisholm}, John and {Pollock}, Clara L. and {Heintz}, Kasper E. and {Johnson}, Benjamin D. and {Shen}, Xuejian and {Hviding}, Raphael E. and {Leja}, Joel and {Tacchella}, Sandro and {Ganguly}, Arpita and {Witten}, Callum and {Atek}, Hakim and {Belli}, Sirio and {Bose}, Sownak and {Bouwens}, Rychard and {Dayal}, Pratika and {Decarli}, Roberto and {de Graaff}, Anna and {Fudamoto}, Yoshinobu and {Giovinazzo}, Emma and {Greene}, Jenny E. and {Illingworth}, Garth and {Inoue}, Akio K. and {Kane}, Sarah G. and {Labbe}, Ivo and {Leonova}, Ecaterina and {Marques-Chaves}, Rui and {Meyer}, Romain A. and {Nelson}, Erica J. and {Roberts-Borsani}, Guido and {Schaerer}, Daniel and {Simcoe}, Robert A. and {Stefanon}, Mauro and {Sugahara}, Yuma and {Toft}, Sune and {van der Wel}, Arjen and {van Dokkum}, Pieter and {Walter}, Fabian and {Watson}, Darach and {Weaver}, John R. and {Whitaker}, Katherine E.},
        title = "{A Cosmic Miracle: A Remarkably Luminous Galaxy at $z_{\rm{spec}}=14.44$ Confirmed with JWST}",
      journal = {arXiv e-prints},
         year = 2025,
        month = may,
          eid = {arXiv:2505.11263},
        pages = {arXiv:2505.11263},
          doi = {10.48550/arXiv.2505.11263},
archivePrefix = {arXiv},
       eprint = {2505.11263},
 primaryClass = {astro-ph.GA},
       adsurl = {https://ui.adsabs.harvard.edu/abs/2025arXiv250511263N}
}

@ARTICLE{Bezanson2024ApJ...974...92B,
       author = {{Bezanson}, Rachel and {Labbe}, Ivo and {Whitaker}, Katherine E. and {Leja}, Joel and {Price}, Sedona H. and {Franx}, Marijn and {Brammer}, Gabriel and {Marchesini}, Danilo and {Zitrin}, Adi and {Wang}, Bingjie and {Weaver}, John R. and {Furtak}, Lukas J. and {Atek}, Hakim and {Coe}, Dan and {Cutler}, Sam E. and {Dayal}, Pratika and {van Dokkum}, Pieter and {Feldmann}, Robert and {F{\"o}rster Schreiber}, Natascha M. and {Fujimoto}, Seiji and {Geha}, Marla and {Glazebrook}, Karl and {de Graaff}, Anna and {Greene}, Jenny E. and {Juneau}, St{\'e}phanie and {Kassin}, Susan and {Kriek}, Mariska and {Khullar}, Gourav and {Maseda}, Michael and {Mowla}, Lamiya A. and {Muzzin}, Adam and {Nanayakkara}, Themiya and {Nelson}, Erica J. and {Oesch}, Pascal A. and {Pacifici}, Camilla and {Pan}, Richard and {Papovich}, Casey and {Setton}, David J. and {Shapley}, Alice E. and {Smit}, Renske and {Stefanon}, Mauro and {Taylor}, Edward N. and {Williams}, Christina C.},
        title = "{The JWST UNCOVER Treasury Survey: Ultradeep NIRSpec and NIRCam Observations before the Epoch of Reionization}",
      journal = {\apj},
         year = 2024,
        month = oct,
       volume = {974},
       number = {1},
          eid = {92},
        pages = {92},
          doi = {10.3847/1538-4357/ad66cf},
archivePrefix = {arXiv},
       eprint = {2212.04026},
 primaryClass = {astro-ph.GA},
       adsurl = {https://ui.adsabs.harvard.edu/abs/2024ApJ...974...92B}
}

@ARTICLE{Fujimoto2024ApJ...977..250F,
       author = {{Fujimoto}, Seiji and {Wang}, Bingjie and {Weaver}, John R. and {Kokorev}, Vasily and {Atek}, Hakim and {Bezanson}, Rachel and {Labbe}, Ivo and {Brammer}, Gabriel and {Greene}, Jenny E. and {Chemerynska}, Iryna and {Dayal}, Pratika and {de Graaff}, Anna and {Furtak}, Lukas J. and {Oesch}, Pascal A. and {Setton}, David J. and {Price}, Sedona H. and {Miller}, Tim B. and {Williams}, Christina C. and {Whitaker}, Katherine E. and {Zitrin}, Adi and {Cutler}, Sam E. and {Leja}, Joel and {Pan}, Richard and {Coe}, Dan and {van Dokkum}, Pieter and {Feldmann}, Robert and {Fudamoto}, Yoshinobu and {Goulding}, Andy D. and {Khullar}, Gourav and {Marchesini}, Danilo and {Maseda}, Michael and {Nanayakkara}, Themiya and {Nelson}, Erica J. and {Smit}, Renske and {Stefanon}, Mauro and {Weibel}, Andrea},
        title = "{UNCOVER: A NIRSpec Census of Lensed Galaxies at z = 8.50{\textendash}13.08 Probing a High-AGN Fraction and Ionized Bubbles in the Shadow}",
      journal = {\apj},
         year = 2024,
        month = dec,
       volume = {977},
       number = {2},
          eid = {250},
        pages = {250},
          doi = {10.3847/1538-4357/ad9027},
archivePrefix = {arXiv},
       eprint = {2308.11609},
 primaryClass = {astro-ph.GA},
       adsurl = {https://ui.adsabs.harvard.edu/abs/2024ApJ...977..250F}
}

@software{bushouse_2025_17515973,
  author       = {Bushouse, Howard and
                  Eisenhamer, Jonathan and
                  Dencheva, Nadia and
                  Davies, James and
                  Greenfield, Perry and
                  Morrison, Jane and
                  Hodge, Phil and
                  Simon, Bernie and
                  Grumm, David and
                  Droettboom, Michael and
                  Slavich, Edward and
                  Sosey, Megan and
                  Pauly, Tyler and
                  Miller, Todd and
                  Jedrzejewski, Robert and
                  Hack, Warren and
                  Davis, David and
                  Crawford, Steven and
                  Law, David and
                  Gordon, Karl and
                  Regan, Michael and
                  Cara, Mihai and
                  MacDonald, Ken and
                  Bradley, Larry and
                  Shanahan, Clare and
                  Jamieson, William and
                  Teodoro, Mairan and
                  Williams, Thomas and
                  Pena-Guerrero, Maria and
                  Graham, Brett and
                  Molter, Edward and
                  Brandt, Timothy and
                  Hayes, Christian and
                  Cooper, Rachel and
                  Clarke, Melanie and
                  Filippazzo, Joseph},
  title        = {JWST Calibration Pipeline},
  month        = nov,
  year         = 2025,
  publisher    = {Zenodo},
  version      = {1.20.2},
  doi          = {10.5281/zenodo.17515973},
  url          = {https://doi.org/10.5281/zenodo.17515973},
  swhid        = {swh:1:dir:fc8e0b17375bd6292e6f2fe6b758a3e44b81aa01
                   ;origin=https://doi.org/10.5281/zenodo.6984365;vis
                   it=swh:1:snp:ee0c72d562544a3226903308be9ab3969e139
                   858;anchor=swh:1:rel:4164a7a6a89b54a00ff5ed5b74c67
                   5e41f26b3fc;path=spacetelescope-jwst-ee52a96
                  },
}

@ARTICLE{Austin2025ApJ...995...43A,
       author = {{Austin}, Duncan and {Conselice}, Christopher J. and {Adams}, Nathan J. and {Harvey}, Thomas and {Duan}, Qiao and {Trussler}, James and {Li}, Qiong and {Juod{\v{z}}balis}, Ignas and {Ormerod}, Katherine and {Ferreira}, Leonardo and {Westcott}, Lewi and {Harris}, Honor and {Wilkins}, Stephen M. and {Bhatawdekar}, Rachana and {Caruana}, Joseph and {Coe}, Dan and {Cohen}, Seth H. and {Driver}, Simon P. and {D'Silva}, Jordan C.~J. and {Frye}, Brenda and {Furtak}, Lukas J. and {Grogin}, Norman A. and {Hathi}, Nimish P. and {Holwerda}, Benne W. and {Jansen}, Rolf A. and {Koekemoer}, Anton M. and {Marshall}, Madeline A. and {Nonino}, Mario and {Ortiz}, III, Rafael and {Pirzkal}, Nor and {Robotham}, Aaron and {Ryan}, Jr., Russell E. and {Summers}, Jake and {Willmer}, Christopher N.~A. and {Windhorst}, Rogier A. and {Yan}, Haojing and {Zackrisson}, Erik},
        title = "{EPOCHS. III. Unbiased UV Continuum Slopes at 6.5 < z < 13 from Combined PEARLS GTO and Public JWST/NIRCam Imaging}",
      journal = {\apj},
         year = 2025,
        month = dec,
       volume = {995},
       number = {1},
          eid = {43},
        pages = {43},
          doi = {10.3847/1538-4357/ae07db},
archivePrefix = {arXiv},
       eprint = {2404.10751},
 primaryClass = {astro-ph.GA},
       adsurl = {https://ui.adsabs.harvard.edu/abs/2025ApJ...995...43A}
}

@ARTICLE{Cullen2024MNRAS.531..997C,
       author = {{Cullen}, F. and {McLeod}, D.~J. and {McLure}, R.~J. and {Dunlop}, J.~S. and {Donnan}, C.~T. and {Carnall}, A.~C. and {Keating}, L.~C. and {Magee}, D. and {Arellano-Cordova}, K.~Z. and {Bowler}, R.~A.~A. and {Begley}, R. and {Flury}, S.~R. and {Hamadouche}, M.~L. and {Stanton}, T.~M.},
        title = "{The ultraviolet continuum slopes of high-redshift galaxies: evidence for the emergence of dust-free stellar populations at z > 10}",
      journal = {\mnras},
         year = 2024,
        month = jun,
       volume = {531},
       number = {1},
        pages = {997-1020},
          doi = {10.1093/mnras/stae1211},
archivePrefix = {arXiv},
       eprint = {2311.06209},
 primaryClass = {astro-ph.GA},
       adsurl = {https://ui.adsabs.harvard.edu/abs/2024MNRAS.531..997C}
}

@ARTICLE{Cullen2023MNRAS.520...14C,
       author = {{Cullen}, Fergus and {McLure}, R.~J. and {McLeod}, D.~J. and {Dunlop}, J.~S. and {Donnan}, C.~T. and {Carnall}, A.~C. and {Bowler}, R.~A.~A. and {Begley}, R. and {Hamadouche}, M.~L. and {Stanton}, T.~M.},
        title = "{The ultraviolet continuum slopes ({\ensuremath{\beta}}) of galaxies at z ≃ 8-16 from JWST and ground-based near-infrared imaging}",
      journal = {\mnras},
         year = 2023,
        month = mar,
       volume = {520},
       number = {1},
        pages = {14-23},
          doi = {10.1093/mnras/stad073},
archivePrefix = {arXiv},
       eprint = {2208.04914},
 primaryClass = {astro-ph.GA},
       adsurl = {https://ui.adsabs.harvard.edu/abs/2023MNRAS.520...14C}
}

@ARTICLE{Finkelstein2019ApJ...879...36F,
       author = {{Finkelstein}, Steven L. and {D'Aloisio}, Anson and {Paardekooper}, Jan-Pieter and {Ryan}, Jr., Russell and {Behroozi}, Peter and {Finlator}, Kristian and {Livermore}, Rachael and {Upton Sanderbeck}, Phoebe R. and {Dalla Vecchia}, Claudio and {Khochfar}, Sadegh},
        title = "{Conditions for Reionizing the Universe with a Low Galaxy Ionizing Photon Escape Fraction}",
      journal = {\apj},
         year = 2019,
        month = jul,
       volume = {879},
       number = {1},
          eid = {36},
        pages = {36},
          doi = {10.3847/1538-4357/ab1ea8},
archivePrefix = {arXiv},
       eprint = {1902.02792},
 primaryClass = {astro-ph.CO},
       adsurl = {https://ui.adsabs.harvard.edu/abs/2019ApJ...879...36F}
}

@ARTICLE{Jaskot2024ApJ...973..111J,
       author = {{Jaskot}, Anne E. and {Silveyra}, Anneliese C. and {Plantinga}, Anna and {Flury}, Sophia R. and {Hayes}, Matthew and {Chisholm}, John and {Heckman}, Timothy and {Pentericci}, Laura and {Schaerer}, Daniel and {Trebitsch}, Maxime and {Verhamme}, Anne and {Carr}, Cody and {Ferguson}, Henry C. and {Ji}, Zhiyuan and {Giavalisco}, Mauro and {Henry}, Alaina and {Marques-Chaves}, Rui and {{\"O}stlin}, G{\"o}ran and {Saldana-Lopez}, Alberto and {Scarlata}, Claudia and {Worseck}, G{\'a}bor and {Xu}, Xinfeng},
        title = "{Multivariate Predictors of Lyman Continuum Escape. II. Predicting Lyman Continuum Escape Fractions for High-redshift Galaxies}",
      journal = {\apj},
         year = 2024,
        month = oct,
       volume = {973},
       number = {2},
          eid = {111},
        pages = {111},
          doi = {10.3847/1538-4357/ad5557},
archivePrefix = {arXiv},
       eprint = {2406.10179},
 primaryClass = {astro-ph.GA},
       adsurl = {https://ui.adsabs.harvard.edu/abs/2024ApJ...973..111J}
}

@ARTICLE{Begley2026MNRAS.545S1995B,
       author = {{Begley}, R. and {McLure}, R.~J. and {Cullen}, F. and {Carnall}, A.~C. and {Stanton}, T.~M. and {Scholte}, D. and {McLeod}, D.~J. and {Dunlop}, J.~S. and {Arellano-C{\'o}rdova}, K.~Z. and {Bondestam}, C. and {Donnan}, C.~T. and {Hamadouche}, M.~L. and {Leung}, H.-H. and {Shapley}, A.~E. and {Stevenson}, S.},
        title = "{The JWST EXCELS survey: A spectroscopic investigation of the ionizing properties of star-forming galaxies at 1<z<8}",
      journal = {\mnras},
         year = 2026,
        month = jan,
       volume = {545},
       number = {1},
          eid = {staf1995},
        pages = {staf1995},
          doi = {10.1093/mnras/staf1995},
archivePrefix = {arXiv},
       eprint = {2509.26591},
 primaryClass = {astro-ph.GA},
       adsurl = {https://ui.adsabs.harvard.edu/abs/2026MNRAS.545S1995B}
}

@ARTICLE{Mascia2024A&A...685A...3M,
       author = {{Mascia}, S. and {Pentericci}, L. and {Calabr{\`o}}, A. and {Santini}, P. and {Napolitano}, L. and {Arrabal Haro}, P. and {Castellano}, M. and {Dickinson}, M. and {Ocvirk}, P. and {Lewis}, J.~S.~W. and {Amor{\'\i}n}, R. and {Bagley}, M. and {Bhatawdekar}, R. and {Cleri}, N.~J. and {Costantin}, L. and {Dekel}, A. and {Finkelstein}, S.~L. and {Fontana}, A. and {Giavalisco}, M. and {Grogin}, N.~A. and {Hathi}, N.~P. and {Hirschmann}, M. and {Holwerda}, B.~W. and {Jung}, I. and {Kartaltepe}, J.~S. and {Koekemoer}, A.~M. and {Lucas}, R.~A. and {Papovich}, C. and {P{\'e}rez-Gonz{\'a}lez}, P.~G. and {Pirzkal}, N. and {Trump}, J.~R. and {Wilkins}, S.~M. and {Yung}, L.~Y.~A.},
        title = "{New insight on the nature of cosmic reionizers from the CEERS survey}",
      journal = {\aap},
         year = 2024,
        month = may,
       volume = {685},
          eid = {A3},
        pages = {A3},
          doi = {10.1051/0004-6361/202347884},
archivePrefix = {arXiv},
       eprint = {2309.02219},
 primaryClass = {astro-ph.GA},
       adsurl = {https://ui.adsabs.harvard.edu/abs/2024A&A...685A...3M}
}

@ARTICLE{Robertson2015ApJ...802L..19R,
       author = {{Robertson}, Brant E. and {Ellis}, Richard S. and {Furlanetto}, Steven R. and {Dunlop}, James S.},
        title = "{Cosmic Reionization and Early Star-forming Galaxies: A Joint Analysis of New Constraints from Planck and the Hubble Space Telescope}",
      journal = {\apjl},
         year = 2015,
        month = apr,
       volume = {802},
       number = {2},
          eid = {L19},
        pages = {L19},
          doi = {10.1088/2041-8205/802/2/L19},
archivePrefix = {arXiv},
       eprint = {1502.02024},
 primaryClass = {astro-ph.CO},
       adsurl = {https://ui.adsabs.harvard.edu/abs/2015ApJ...802L..19R}
}

@ARTICLE{Pahl2025ApJ...981..134P,
       author = {{Pahl}, Anthony and {Topping}, Michael W. and {Shapley}, Alice and {Sanders}, Ryan and {Reddy}, Naveen A. and {Clarke}, Leonardo and {Kehoe}, Emily and {Bento}, Trinity and {Brammer}, Gabe},
        title = "{A Spectroscopic Analysis of the Ionizing Photon Production Efficiency in JADES and CEERS: Implications for the Ionizing Photon Budget}",
      journal = {\apj},
         year = 2025,
        month = mar,
       volume = {981},
       number = {2},
          eid = {134},
        pages = {134},
          doi = {10.3847/1538-4357/adb1ab},
archivePrefix = {arXiv},
       eprint = {2407.03399},
 primaryClass = {astro-ph.GA},
       adsurl = {https://ui.adsabs.harvard.edu/abs/2025ApJ...981..134P}
}

@ARTICLE{Kendrew2015PASP..127..623K,
       author = {{Kendrew}, Sarah and {Scheithauer}, Silvia and {Bouchet}, Patrice and {Amiaux}, Jerome and {Azzollini}, Ruym{\'a}n and {Bouwman}, Jeroen and {Chen}, C.~H. and {Dubreuil}, D. and {Fischer}, Sebastian and {Glasse}, Alistair and {Greene}, T.~P. and {Lagage}, P.-O. and {Lahuis}, Fred and {Ronayette}, Samuel and {Wright}, David and {Wright}, G.~S.},
        title = "{The Mid-Infrared Instrument for the James Webb Space Telescope, IV: The Low-Resolution Spectrometer}",
      journal = {\pasp},
         year = 2015,
        month = jul,
       volume = {127},
       number = {953},
        pages = {623},
          doi = {10.1086/682255},
archivePrefix = {arXiv},
       eprint = {1512.03000},
 primaryClass = {astro-ph.IM},
       adsurl = {https://ui.adsabs.harvard.edu/abs/2015PASP..127..623K}
}

@ARTICLE{Hainline2024ApJ...976..160H,
       author = {{Hainline}, Kevin N. and {D'Eugenio}, Francesco and {Jakobsen}, Peter and {Chevallard}, Jacopo and {Carniani}, Stefano and {Witstok}, Joris and {Ji}, Zhiyuan and {Curtis-Lake}, Emma and {Johnson}, Benjamin D. and {Robertson}, Brant and {Tacchella}, Sandro and {Curti}, Mirko and {Charlot}, Stephane and {Helton}, Jakob M. and {Arribas}, Santiago and {Bhatawdekar}, Rachana and {Bunker}, Andrew J. and {Cameron}, Alex J. and {Egami}, Eiichi and {Eisenstein}, Daniel J. and {Hausen}, Ryan and {Kumari}, Nimisha and {Maiolino}, Roberto and {P{\'e}rez-Gonz{\'a}lez}, Pablo G. and {Rieke}, Marcia and {Saxena}, Aayush and {Scholtz}, Jan and {Smit}, Renske and {Sun}, Fengwu and {Williams}, Christina C. and {Willmer}, Christopher N.~A. and {Willott}, Chris},
        title = "{Searching for Emission Lines at z > 11: The Role of Damped Ly{\ensuremath{\alpha}} and Hints About the Escape of Ionizing Photons}",
      journal = {\apj},
         year = 2024,
        month = dec,
       volume = {976},
       number = {2},
          eid = {160},
        pages = {160},
          doi = {10.3847/1538-4357/ad8447},
archivePrefix = {arXiv},
       eprint = {2404.04325},
 primaryClass = {astro-ph.GA},
       adsurl = {https://ui.adsabs.harvard.edu/abs/2024ApJ...976..160H}
}

@ARTICLE{Dessauges-Zavadsky2025A&A...693A..17D,
       author = {{Dessauges-Zavadsky}, M. and {Marques-Chaves}, R. and {Schaerer}, D. and {Xiao}, M.-Y. and {Colina}, L. and {Alvarez-Marquez}, J. and {P{\'e}rez-Fournon}, I.},
        title = "{Unveiling dust, molecular gas, and high star-formation efficiency in extremely UV bright star-forming galaxies at z {\ensuremath{\sim}} 2.1{\textendash}3.6}",
      journal = {\aap},
         year = 2025,
        month = jan,
       volume = {693},
          eid = {A17},
        pages = {A17},
          doi = {10.1051/0004-6361/202451832},
archivePrefix = {arXiv},
       eprint = {2410.11121},
 primaryClass = {astro-ph.GA},
       adsurl = {https://ui.adsabs.harvard.edu/abs/2025A&A...693A..17D}
}

@ARTICLE{Komarova2025ApJ...994..192K,
       author = {{Komarova}, Lena and {Oey}, M.~S. and {Marques-Chaves}, Rui and {Amor{\'\i}n}, Ricardo and {Henry}, Alaina and {Schaerer}, Daniel and {Saldana-Lopez}, Alberto and {Le Reste}, Alexandra and {Scarlata}, Claudia and {Hayes}, Matthew J. and {Bait}, Omkar and {Borthakur}, Sanchayeeta and {Carr}, Cody and {Chisholm}, John and {Ferguson}, Harry C. and {Gutierrez Fernandez}, Vital and {Fleming}, Brian and {Flury}, Sophia R. and {Giavalisco}, Mauro and {Grazian}, Andrea and {Heckman}, Timothy and {Jaskot}, Anne E. and {Ji}, Zhiyuan and {{\"O}stlin}, G{\"o}ran and {Pentericci}, Laura and {Ravindranath}, Swara and {Thuan}, Trinh and {V{\'\i}lchez}, Jose M. and {Worseck}, Gabor and {Xu}, Xinfeng},
        title = "{Power-law Emission-line Wings and Radiation-driven Superwinds in Local Lyman Continuum Emitters}",
      journal = {\apj},
         year = 2025,
        month = dec,
       volume = {994},
       number = {2},
          eid = {192},
        pages = {192},
          doi = {10.3847/1538-4357/ae0e0a},
archivePrefix = {arXiv},
       eprint = {2506.19623},
 primaryClass = {astro-ph.GA},
       adsurl = {https://ui.adsabs.harvard.edu/abs/2025ApJ...994..192K}
}

@ARTICLE{Marques-Chaves2025arXiv251012411M,
       author = {{Marques-Chaves}, R. and {Schaerer}, D. and {Dessauges-Zavadsky}, M. and {{\'A}lvarez-M{\'a}rquez}, J. and {Hashimoto}, T. and {Colina}, L. and {Inoue}, A.~K. and {Blanco-Prieto}, C. and {Nakazato}, Y. and {Costantin}, L. and {Arribas}, S. and {Bakx}, T.~J.~L.~C. and {Ceverino}, D. and {Crespo G{\'o}mez}, A. and {Fudamoto}, Y. and {Hagimoto}, M. and {Hamada}, A. and {Matsuoka}, Y. and {Mawatari}, K. and {Onoue}, M. and {Osone}, W. and {Ren}, Y.~W. and {Sugahara}, Y. and {Terui}, Y. and {Yoshida}, N.},
        title = "{Extremely UV-bright starbursts at the end of cosmic reionization}",
      journal = {arXiv e-prints},
         year = 2025,
        month = oct,
          eid = {arXiv:2510.12411},
        pages = {arXiv:2510.12411},
          doi = {10.48550/arXiv.2510.12411},
archivePrefix = {arXiv},
       eprint = {2510.12411},
 primaryClass = {astro-ph.GA},
       adsurl = {https://ui.adsabs.harvard.edu/abs/2025arXiv251012411M}
}

@ARTICLE{Crespo2025arXiv251114658C,
       author = {{Crespo G{\'o}mez}, A. and {Tamura}, Y. and {Colina}, L. and {{\'A}lvarez-M{\'a}rquez}, J. and {Hashimoto}, T. and {Marques-Chaves}, R. and {Nakazato}, Y. and {Blanco-Prieto}, C. and {Sunaga}, K. and {Costantin}, L. and {Inoue}, A.~K. and {Hamada}, A. and {Arribas}, S. and {Ceverino}, D. and {Hagimoto}, M. and {Mawatari}, K. and {Osone}, W. and {Sugahara}, Y. and {Harikane}, Y. and {Lee}, M.~M. and {Taniguchi}, A. and {Umehata}, H.},
        title = "{RIOJA. Dusty outflows and density-complex ISM in the N-enhanced lensed galaxy RXCJ2248-ID at z=6.1}",
      journal = {arXiv e-prints},
         year = 2025,
        month = nov,
          eid = {arXiv:2511.14658},
        pages = {arXiv:2511.14658},
          doi = {10.48550/arXiv.2511.14658},
archivePrefix = {arXiv},
       eprint = {2511.14658},
 primaryClass = {astro-ph.GA},
       adsurl = {https://ui.adsabs.harvard.edu/abs/2025arXiv251114658C}
}

@ARTICLE{Sharma2017MNRAS.468.2176S,
       author = {{Sharma}, Mahavir and {Theuns}, Tom and {Frenk}, Carlos and {Bower}, Richard G. and {Crain}, Robert A. and {Schaller}, Matthieu and {Schaye}, Joop},
        title = "{Winds of change: reionization by starburst galaxies}",
      journal = {\mnras},
         year = 2017,
        month = jun,
       volume = {468},
       number = {2},
        pages = {2176-2188},
          doi = {10.1093/mnras/stx578},
archivePrefix = {arXiv},
       eprint = {1606.08688},
 primaryClass = {astro-ph.CO},
       adsurl = {https://ui.adsabs.harvard.edu/abs/2017MNRAS.468.2176S}
}

@ARTICLE{Naidu2020ApJ...892..109N,
       author = {{Naidu}, Rohan P. and {Tacchella}, Sandro and {Mason}, Charlotte A. and {Bose}, Sownak and {Oesch}, Pascal A. and {Conroy}, Charlie},
        title = "{Rapid Reionization by the Oligarchs: The Case for Massive, UV-bright, Star-forming Galaxies with High Escape Fractions}",
      journal = {\apj},
         year = 2020,
        month = apr,
       volume = {892},
       number = {2},
          eid = {109},
        pages = {109},
          doi = {10.3847/1538-4357/ab7cc9},
archivePrefix = {arXiv},
       eprint = {1907.13130},
 primaryClass = {astro-ph.GA},
       adsurl = {https://ui.adsabs.harvard.edu/abs/2020ApJ...892..109N}
}

@ARTICLE{Rivera-Thorsen2019Sci...366..738R,
       author = {{Rivera-Thorsen}, T. Emil and {Dahle}, H{\r{a}}kon and {Chisholm}, John and {Florian}, Michael K. and {Gronke}, Max and {Rigby}, Jane R. and {Gladders}, Michael D. and {Mahler}, Guillaume and {Sharon}, Keren and {Bayliss}, Matthew},
        title = "{Gravitational lensing reveals ionizing ultraviolet photons escaping from a distant galaxy}",
      journal = {Science},
         year = 2019,
        month = nov,
       volume = {366},
       number = {6466},
        pages = {738-741},
          doi = {10.1126/science.aaw0978},
archivePrefix = {arXiv},
       eprint = {1904.08186},
 primaryClass = {astro-ph.GA},
       adsurl = {https://ui.adsabs.harvard.edu/abs/2019Sci...366..738R}
}

@ARTICLE{Vanzella2016ApJ...825...41V,
       author = {{Vanzella}, E. and {de Barros}, S. and {Vasei}, K. and {Alavi}, A. and {Giavalisco}, M. and {Siana}, B. and {Grazian}, A. and {Hasinger}, G. and {Suh}, H. and {Cappelluti}, N. and {Vito}, F. and {Amorin}, R. and {Balestra}, I. and {Brusa}, M. and {Calura}, F. and {Castellano}, M. and {Comastri}, A. and {Fontana}, A. and {Gilli}, R. and {Mignoli}, M. and {Pentericci}, L. and {Vignali}, C. and {Zamorani}, G.},
        title = "{Hubble Imaging of the Ionizing Radiation from a Star-forming Galaxy at Z=3.2 with fesc>50\%}",
      journal = {\apj},
         year = 2016,
        month = jul,
       volume = {825},
       number = {1},
          eid = {41},
        pages = {41},
          doi = {10.3847/0004-637X/825/1/41},
archivePrefix = {arXiv},
       eprint = {1602.00688},
 primaryClass = {astro-ph.GA},
       adsurl = {https://ui.adsabs.harvard.edu/abs/2016ApJ...825...41V}
}

@ARTICLE{deBarros2016A&A...585A..51D,
       author = {{de Barros}, S. and {Vanzella}, E. and {Amor{\'\i}n}, R. and {Castellano}, M. and {Siana}, B. and {Grazian}, A. and {Suh}, H. and {Balestra}, I. and {Vignali}, C. and {Verhamme}, A. and {Zamorani}, G. and {Mignoli}, M. and {Hasinger}, G. and {Comastri}, A. and {Pentericci}, L. and {P{\'e}rez-Montero}, E. and {Fontana}, A. and {Giavalisco}, M. and {Gilli}, R.},
        title = "{An extreme [O III] emitter at z = 3.2: a low metallicity Lyman continuum source}",
      journal = {\aap},
         year = 2016,
        month = jan,
       volume = {585},
          eid = {A51},
        pages = {A51},
          doi = {10.1051/0004-6361/201527046},
archivePrefix = {arXiv},
       eprint = {1507.06648},
 primaryClass = {astro-ph.GA},
       adsurl = {https://ui.adsabs.harvard.edu/abs/2016A&A...585A..51D}
}

@ARTICLE{Izotov2025A&A...704A..19I,
       author = {{Izotov}, Y.~I. and {Schaerer}, D. and {Worseck}, G. and {Guseva}, N.~G. and {Verhamme}, A. and {Simmonds}, C. and {Chisholm}, J.},
        title = "{A great diversity of spectral shapes in the ionising spectra of z {\ensuremath{\sim}} 0.6─1 galaxies revealed by HST/COS and possible detection of nebular LyC emission}",
      journal = {\aap},
         year = 2025,
        month = nov,
       volume = {704},
          eid = {A19},
        pages = {A19},
          doi = {10.1051/0004-6361/202556004},
archivePrefix = {arXiv},
       eprint = {2510.22152},
 primaryClass = {astro-ph.GA},
       adsurl = {https://ui.adsabs.harvard.edu/abs/2025A&A...704A..19I}
}

@ARTICLE{Simmonds2024MNRAS.530.2133S,
       author = {{Simmonds}, C. and {Verhamme}, A. and {Inoue}, A.~K. and {Katz}, H. and {Garel}, T. and {De Barros}, S.},
        title = "{The impact of nebular Lyman-Continuum on ionizing photons budget and escape fractions from galaxies}",
      journal = {\mnras},
         year = 2024,
        month = may,
       volume = {530},
       number = {2},
        pages = {2133-2145},
          doi = {10.1093/mnras/stae1003},
archivePrefix = {arXiv},
       eprint = {2402.04052},
 primaryClass = {astro-ph.GA},
       adsurl = {https://ui.adsabs.harvard.edu/abs/2024MNRAS.530.2133S}
}

@ARTICLE{Inoue2010MNRAS.401.1325I,
       author = {{Inoue}, Akio K.},
        title = "{Lyman `bump' galaxies - I. Spectral energy distribution of galaxies with an escape of nebular Lyman continuum}",
      journal = {\mnras},
         year = 2010,
        month = jan,
       volume = {401},
       number = {2},
        pages = {1325-1333},
          doi = {10.1111/j.1365-2966.2009.15730.x},
archivePrefix = {arXiv},
       eprint = {0908.3925},
 primaryClass = {astro-ph.CO},
       adsurl = {https://ui.adsabs.harvard.edu/abs/2010MNRAS.401.1325I}
}

@ARTICLE{Kim2023ApJ...955L..17K,
       author = {{Kim}, Keunho J. and {Bayliss}, Matthew B. and {Rigby}, Jane R. and {Gladders}, Michael D. and {Chisholm}, John and {Sharon}, Keren and {Dahle}, H{\r{a}}kon and {Rivera-Thorsen}, T. Emil and {Florian}, Michael K. and {Khullar}, Gourav and {Mahler}, Guillaume and {Mainali}, Ramesh and {Napier}, Kate A. and {Navarre}, Alexander and {Owens}, M. Riley and {Roberson}, Joshua},
        title = "{Small Region, Big Impact: Highly Anisotropic Lyman-continuum Escape from a Compact Starburst Region with Extreme Physical Properties}",
      journal = {\apjl},
         year = 2023,
        month = sep,
       volume = {955},
       number = {1},
          eid = {L17},
        pages = {L17},
          doi = {10.3847/2041-8213/acf0c5},
archivePrefix = {arXiv},
       eprint = {2305.13405},
 primaryClass = {astro-ph.GA},
       adsurl = {https://ui.adsabs.harvard.edu/abs/2023ApJ...955L..17K}
}

@BOOK{Osterbrock+Ferland+06,
       author = {{Osterbrock}, Donald E. and {Ferland}, Gary J.},
        title = "{Astrophysics of gaseous nebulae and active galactic nuclei}",
         year = 2006,
       adsurl = {https://ui.adsabs.harvard.edu/abs/2006agna.book.....O}
}

\onecolumn

\begin{appendix}

\section{BPASS predictions for different metallicities}\label{appendix1}

\begin{figure*}[h!]
\centering
  \includegraphics[width=0.98\textwidth]{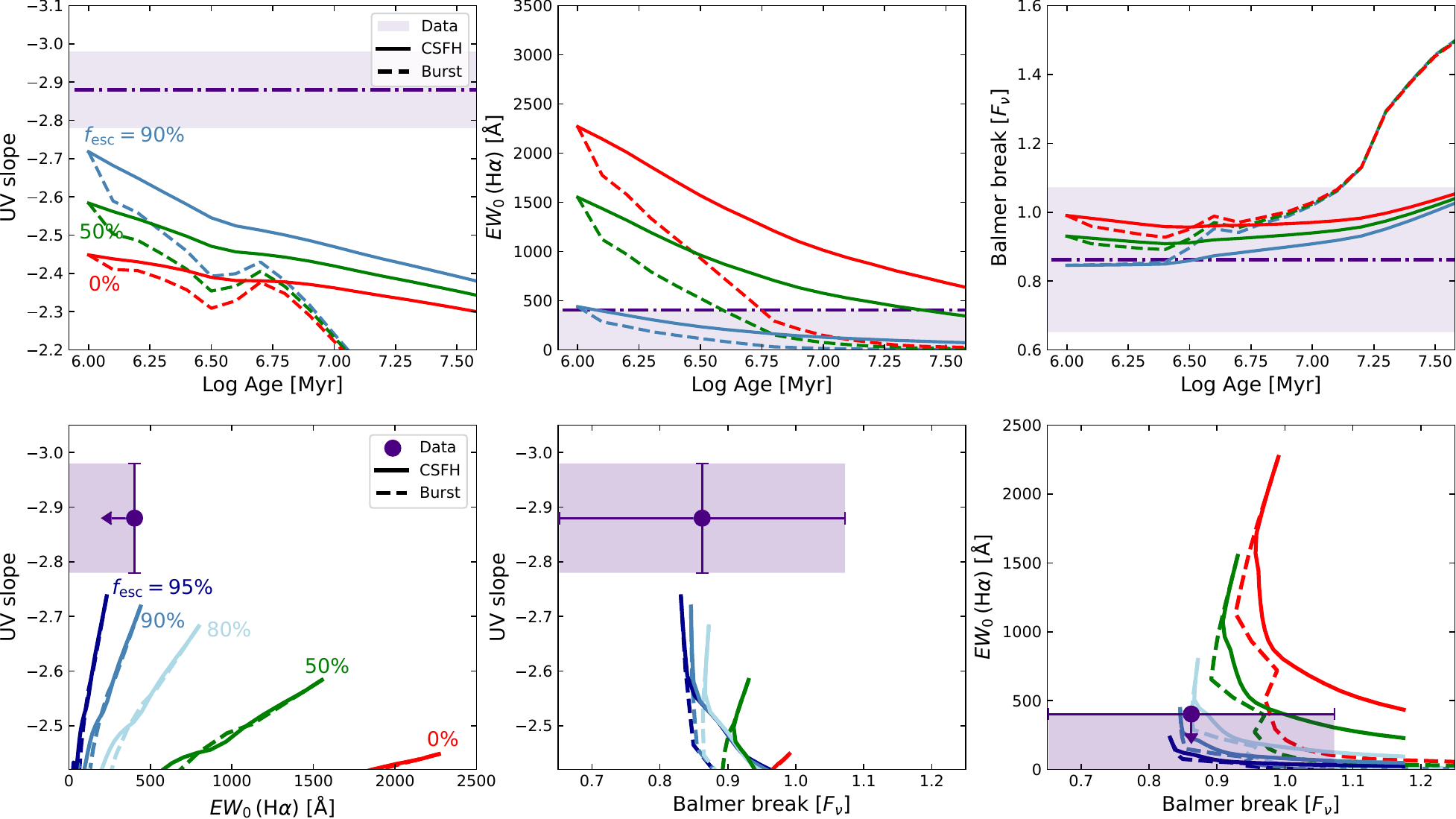}
  \caption{Same as Figure~\ref{fig_BPASS_predictions} but for BPASS models with stellar metallicity of $Z=0.02$ (or $Z/Z_{\odot}=100\%$).}
  \label{fig_appendix_A1}
\end{figure*}

\begin{figure*}[h!]
\centering
  \includegraphics[width=0.98\textwidth]{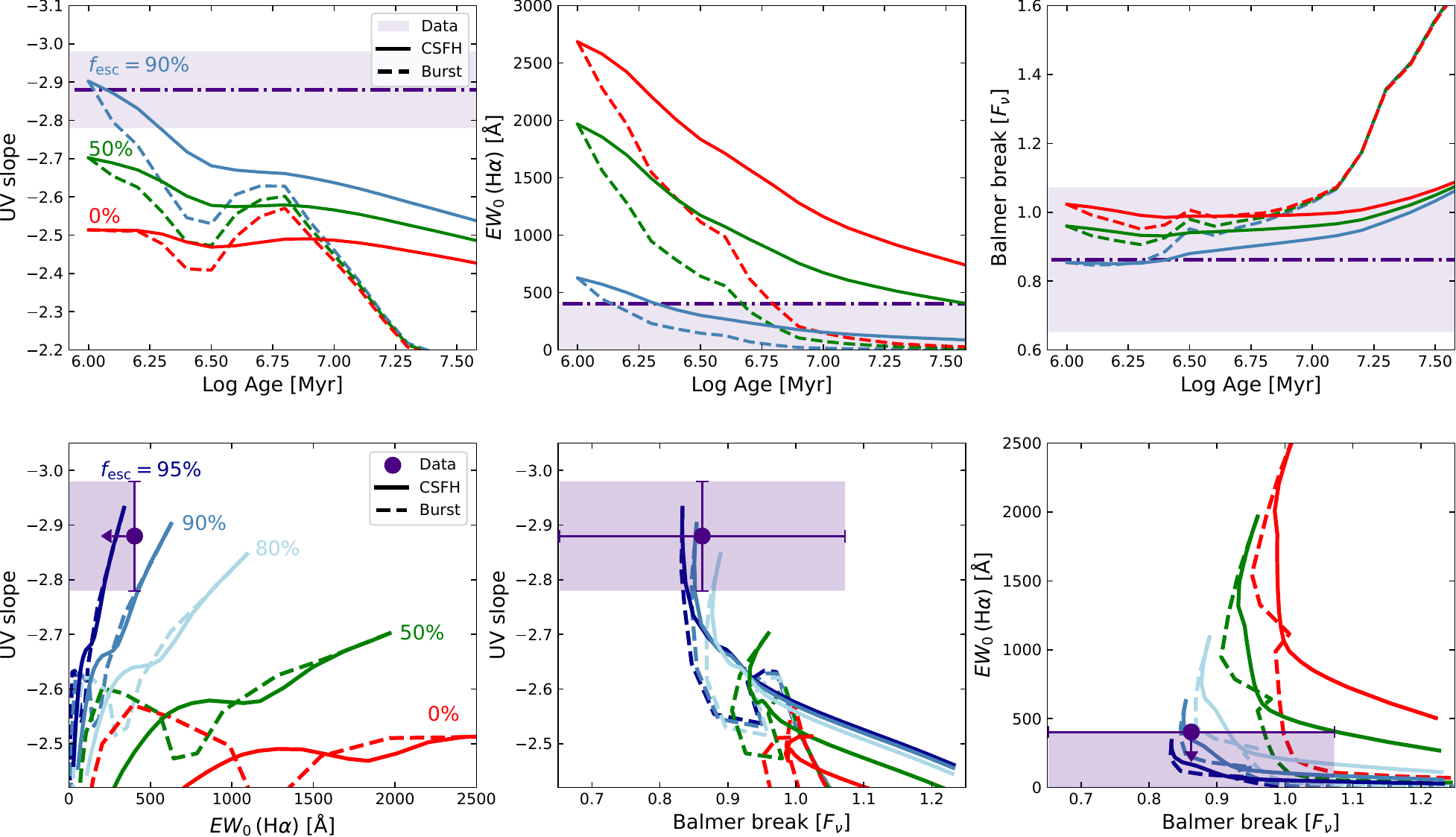}
  \caption{Same as Figure~\ref{fig_BPASS_predictions} but for BPASS models with stellar metallicity of $Z=0.01$ (or $Z/Z_{\odot}=50\%$).}
  \label{fig_appendix_A2}
\end{figure*}

\begin{figure*}[h!]
\centering
  \includegraphics[width=0.98\textwidth]{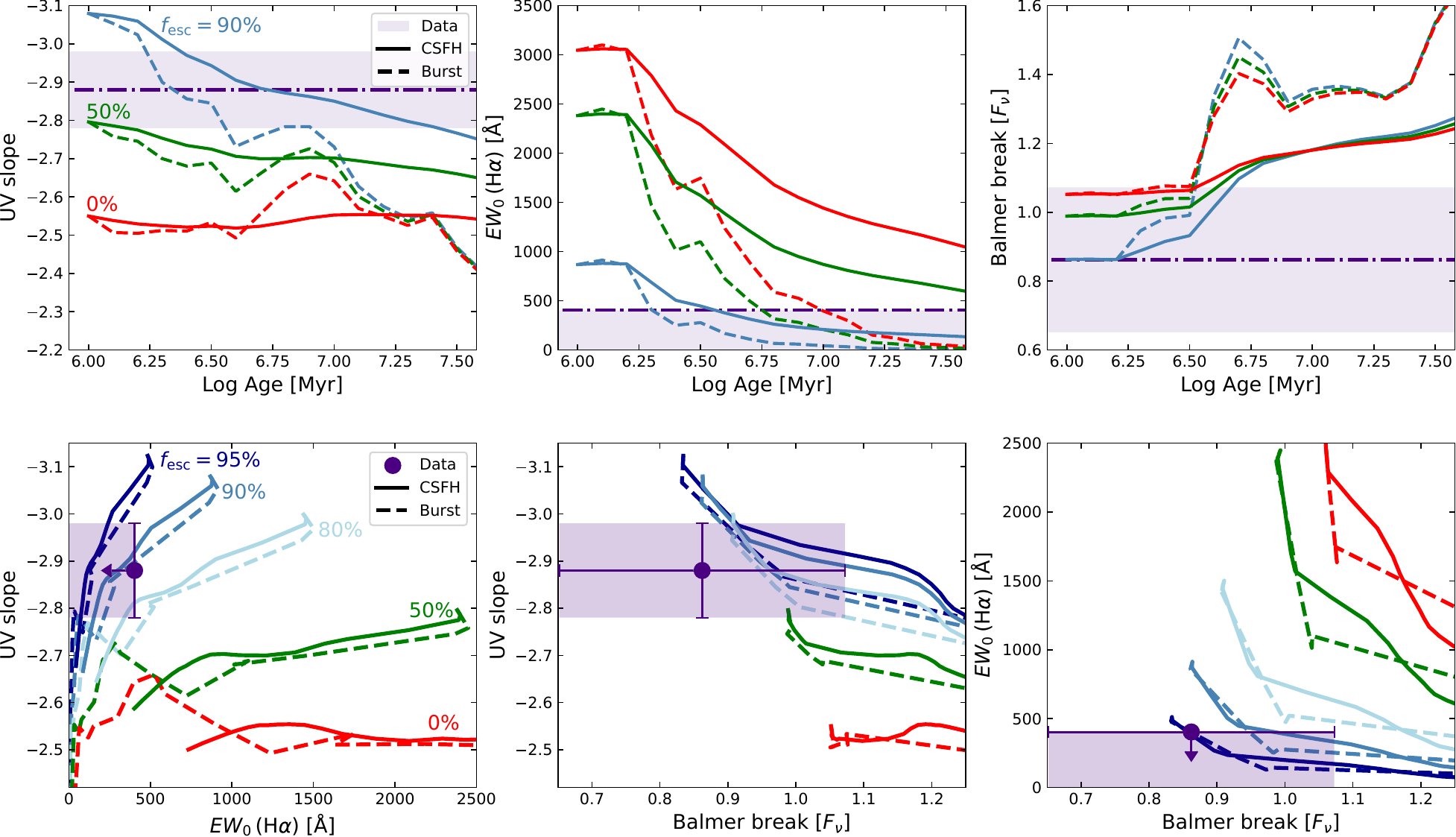}
  \caption{Same as Figure~\ref{fig_BPASS_predictions} but for BPASS models with stellar metallicity of $Z=0.001$ (or $Z/Z_{\odot}=5\%$).}
  \label{fig_appendix_A3}
\end{figure*}

\begin{figure*}[h!]
\centering
  \includegraphics[width=0.98\textwidth]{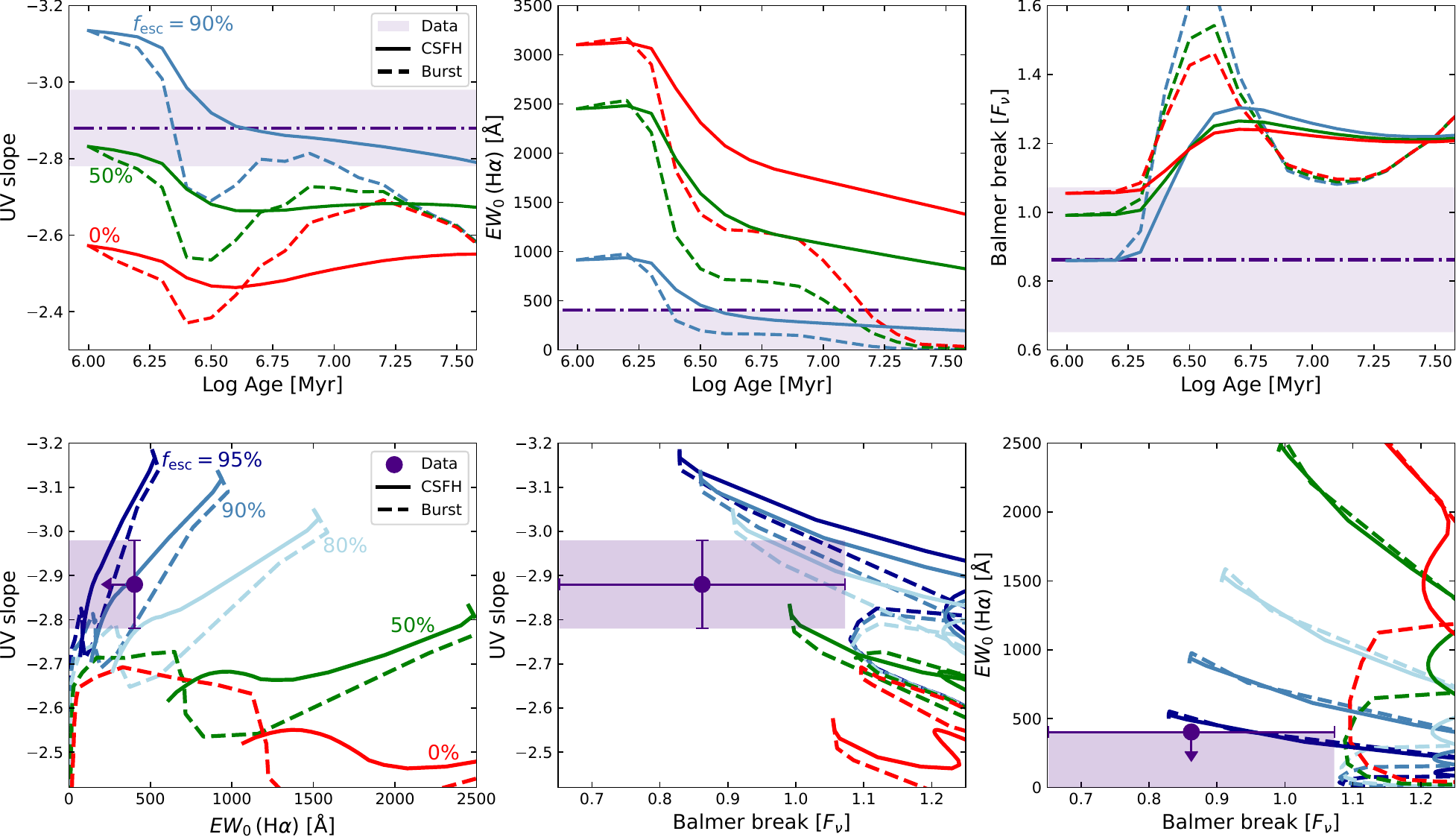}
  \caption{Same as Figure~\ref{fig_BPASS_predictions} but for BPASS models with stellar metallicity of $Z=0.0001$ (or $Z/Z_{\odot}=0.5\%$).}
  \label{fig_appendix_A4}
\end{figure*}

Figures~\ref{fig_appendix_A1}, \ref{fig_appendix_A2}, \ref{fig_appendix_A3}, and \ref{fig_appendix_A4} show the model predictions of the UV slope, $EW_{0}$\,(H$\alpha$) and the Balmer break strength as a function of age (top) and their combination (bottom) for stellar metallicities of $Z=Z_{\odot}$, $Z/Z_{\odot}=50\%$, $Z/Z_{\odot}=5\%$, and $Z/Z_{\odot}=0.5\%$, respectively.

\end{appendix}

\end{document}